\documentclass[10pt,conference,letterpaper]{IEEEtran}
\usepackage{times,amsmath,epsfig}

\usepackage{amsfonts}
\usepackage[ruled,linesnumbered,noend,noline]{algorithm2e}
\usepackage{xspace}
\usepackage{subfigure}
\usepackage{url}
\usepackage{threeparttable}

\usepackage[noadjust]{cite}

\usepackage{tikz}
\def\checkmark{\tikz\fill[scale=0.4](0,.35) -- (.25,0) -- (1,.7) -- (.25,.15) -- cycle;} 
\title{CiNCT: Compression and retrieval for massive vehicular trajectories via relative movement labeling}
\author{%
{Satoshi Koide{\small $~^{\#*1}$}, Yukihiro Tadokoro{\small $~^{\#2}$}, Chuan Xiao{\small $~^{*3}$}, Yoshiharu Ishikawa{\small $~^{*4}$} }%
\vspace{1.6mm}\\
\fontsize{10}{10}\selectfont\itshape
$^{\#}$\,Toyota Central R\&D Labs., Inc., Nagakute, Aichi, Japan\\
$^{*}$\,Nagoya University, Nagoya, Aichi, Japan\\[.2em]
\fontsize{9}{9}\selectfont\ttfamily\upshape
$^{1}$\,koide@mosk.tytlabs.co.jp
$^{2}$\,tadokoro@mosk.tytlabs.co.jp\\%
$^{3}$\,chuanx@nagoya-u.jp
$^{4}$\,ishikawa@i.nagoya-u.ac.jp
}
\newcommand{\LandC}{MEL\xspace}
\newcommand{\nonl}{\renewcommand{\nl}{\let\nl\oldnl}}
\newcommand{\proposedbase}{CiNCT}
\newcommand{\proposed}{\proposedbase\xspace}
\newcommand{\proposedsf}{\textsf{\proposedbase}\xspace}
\newcommand{\bnd}{\textsf{\$}}
\newcommand{\bsharp}{\textsf{\#}}
\newcommand{\mypara}{\subsubsection}

\newtheorem{defi}{Definition}
\newtheorem{theo}{Theorem}

\newtheorem{lemm}{Lemma}

\begin{document}
\maketitle
\begin{abstract}
In this paper, we present a compressed data structure for moving object trajectories in a road network, which are represented as sequences of road edges.
Unlike existing compression methods for trajectories in a network, our method supports pattern matching and decompression from an arbitrary position while retaining a high compressibility with theoretical guarantees.
Specifically, our method is based on FM-index, a fast and compact data structure for pattern matching.
To enhance the compression, we incorporate the sparsity of road networks into the data structure.
In particular, we present the novel concepts of \emph{relative movement labeling} and \emph{PseudoRank}, each contributing to significant reductions in data size and query processing time.
Our theoretical analysis and experimental studies reveal the advantages of our proposed method as compared to existing trajectory compression methods and FM-index variants.
%
\end{abstract}

%

\section{Introduction}
\label{sec:intro}
In recent years, a vast amount of trajectory data from moving objects, such as automobiles, has become available.
According to Han et al.~\cite{COMPRESS}, the total amount of GPS trajectories generated by automobiles in the U.S. alone exceeded 53~TB in 2011.
With recent increased interest in the use of such large datasets in wide range of data-driven applications, fundamental data manipulations such as retrieval and compression are once again becoming crucial.
In this paper, we focus on moving object trajectories in (road) networks, called \emph{network-constrained trajectories} (NCTs), one of the most important types of trajectories with many practical applications.
Traveled paths of NCTs can be represented as symbol sequences of road segment IDs.
Although this representation is more compact than GPS coordinates, it is still insufficient for the vast datasets that are now available.
Therefore, compressed representations of NCTs have been studied thus far \cite{COMPRESS, MMTC, CTR, SPNET, PopaSTC}.

If trajectories are simply compressed without an augmented data structure, it is difficult to use them in real applications.
Therefore, compression methods that allow several operations without decompressing the entire dataset are necessary, and such methods have been the focus of recent studies.
For example, such studies include the in-memory data structures proposed in \cite{CTR} and \cite{SPNET}, as well as an in-memory/on-disk hybrid structure proposed in \cite{SNT-index}.
In our present paper, we propose a method that realizes a high level of compression while retaining a high utility of the data.
As motivation and background for our method, we first review existing compressed data structures for NCTs and their functions below.

NCTs consist of spatial paths and corresponding timestamps.
We therefore must consider compression of these paths and timestamps separately.
For spatial paths, lossless compression methods based on \emph{shortest-path encoding} have been studied in \cite{COMPRESS, MMTC}, and \cite{SPNET}.
Here, to compress the data, these methods remove partial shortest paths in an NCT because these paths can be recovered from the road network itself.
One drawback of this approach is that it cannot guarantee the information-theoretic upper bound of the compressed data size.
A recent lossless path compressor introduced in \cite{COMPRESS} called \emph{minimum entropy labeling} (\LandC) guarantees a theoretic bound and also achieves practically higher compressibility than shortest-path encoding methods.
As for the timestamps, all methods noted above compress them independently from the spatial path compression.
In this paper, we do not discuss the compression of timestamps directly, but we emphasize here that our method can be easily combined with such temporal compression methods (see Section \ref{sec:related-work} for details).

In general, it is difficult to define high utility of compressed NCTs, because their utility depends on the given application.
In this paper, we focus on two functions, i.e., pattern matching without decompressing the entire dataset, and extracting sub-paths from an arbitrary position.
Intuitively, pattern matching operations that find trajectories along a given path would have wide applications in NCT processing.
In fact, the existing methods mentioned above (i.e., \cite{CTR, SPNET}, and \cite{SNT-index}) closely relate to pattern matching; however, to the best of our knowledge, 
there are no NCT compressors that guarantee theoretical bound for the compressed size while supporting fast pattern matching.

Given the above, our research question is, \emph{how can we realize high compressibility while enabling pattern matching for NCTs?}
To address this question, we focus on \emph{suffix arrays} \cite{SA}, data structures that closely relate to pattern matching.
Although the data structures for NCTs proposed in \cite{CTR} and \cite{SNT-index} also employ suffix arrays, they do not focus on a compression method, instead using existing general-purpose compressed suffix arrays that are typically used for handling genomic sequences.
Unfortunately, these existing methods are inefficient because genomic sequences include only four characters (i.e, A, C, G, and T) whereas NCTs consist of a large alphabet (i.e., road segment IDs in a potentially large road network).

\begin{figure}[tbh]
 \centering
 \includegraphics[scale=.95]{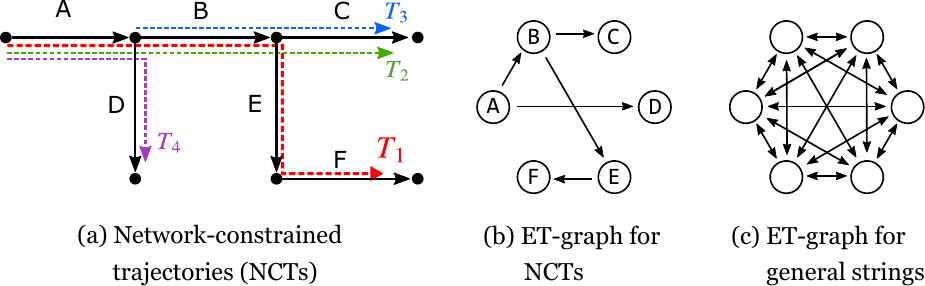}
 \caption{(a) Network-constrained trajectories (NCTs), and both (b) sparse and (c) dense symbol transition graphs (ET-graphs).}
 \label{fig:main-idea}
\end{figure}

NCTs have another noteworthy feature, i.e., \emph{they can only move along physically connected road segments}.
This feature is quite different from general sequences, as illustrated in Fig.~\ref{fig:main-idea}.
In Fig.~\ref{fig:main-idea}(a), we show four example NCTs in a small network with six road segments (\textsf{A}--\textsf{F}).
The corresponding graph shown in Fig.~\ref{fig:main-idea}(b) represents symbol transitions for these four NCTs.
Here, each vertex corresponds to a symbol (i.e., a road segment), and directed edges exist between two vertices if the corresponding two symbols can appear successively.
For example, in Fig.~\ref{fig:main-idea}(b), vertex \textsf{A} is connected with vertexes \textsf{B} and \textsf{D} because we can only move to road segment \textsf{B} or \textsf{D} from \textsf{A}.
For NCTs, this \emph{empirical transition graph} (ET-graph) becomes a sparse graph, reflecting the physical topology of road networks. 
This sparsity cannot be obtained for general sequences, which leads to a denser ET-graph, as illustrated in Fig.~\ref{fig:main-idea}(c).

Our proposed method, \emph{Compressed-index for NCTs} (\emph{\proposedsf}), significantly improves the compression and pattern matching operations when applied to sequences with such sparse ET-graphs.
Our method is based on \emph{FM-index} \cite{FM-index}, a compressed data structure for suffix arrays, which we describe further in Section \ref{sec:preliminaries}.
Note that it is challenging to incorporate such sparsity into FM-index while retaining its theoretical advantages because FM-index is compressed at the bit-level.
Therefore, in the remainder of our paper, we introduce some novel techniques and provide theoretical analysis that explains why our method yields substantial improvement in practice.

\textit{Contributions}:
To develop a data structure for NCTs that simultaneously achieves a high compression ratio and high utility, we propose \proposedsf, as a novel method to compress suffix arrays for sequences on a sparse graph.
We summarize our contributions as follows.
\begin{itemize}
 \item We propose \emph{relative movement labeling} (RML), which converts sequences on a sparse graph to low-entropy sequences. We theoretically prove its optimality and show that RML provides a more compact representation of NCTs than that of the \LandC method \cite{COMPRESS}.
 \item We incorporate RML into FM-index by introducing a new concept called \emph{PseudoRank}, which leads to significant improvements in both size and query processing speed (i.e., the speed of pattern matching and sub-path extraction) as compared to existing FM-index variants. We also explain theoretically why this occurs.
 \item Using several real NCT datasets, we show that our method outperforms state-of-the-art methods that do not consider graph sparsity.
\end{itemize}
\textit{Outline}: The remainder of our paper is organized as follows:
preliminaries (Section~\ref{sec:preliminaries}),
proposed data structure (Section~\ref{sec:proposed}),
proposed algorithms (Section~\ref{sec:search}),
theoretical analysis (Section~\ref{sec:theory}),
experiments (Section~\ref{sec:experiment}),
related work (Section~\ref{sec:related-work}),
and conclusion (Section~\ref{sec:conclusion}).

\section{Preliminaries}
\label{sec:preliminaries}
In this section, we introduce the data models and pattern matching query. 
For readers not familiar with string processing and indexing, we also describe the necessary concepts regarding FM-index and its compression.
Table \ref{table:notation} summarizes notation used in this paper.
\begin{table}[t]
 \centering
 \caption{Notation}\label{table:notation}
 \vspace{-.8em}
 \begin{tabular}{lll} \hline
  \textbf{Symbol} & \textbf{Description} & \textbf{Defined in} \\ \hline
  $w, w'\in E$ & Road segments (characters)    & ---  \\
  $T, T_{bwt}$          & Trajectory string and its BWT& Def.~\ref{def:trajstr}, Fig.~\ref{fig:bwt}\\
  $\Sigma, \sigma$ & Alphabet set and its size & \S~\ref{sssec:trajstr}\\
  $R(P)\!=\![sp,ep)$       & Suffix range of a pattern $P$ & \S~\ref{sssec:bwt}\\
  $C[w]$       & The number of $w'$ in $T$ s.t.~$w'\!<\!w$ & \S~\ref{sssec:fmalgo}\\
  $H_0(S)$, $H_k(S)$ & 0th and $k$th order empirical entropy& Eq.~(\ref{eq:H0}), (\ref{eq:Hk})\\ 
  $G_T, E_T$   & ET-graph and its edge set & \S~\ref{sec:traj encoding} (Def.~\ref{def:ET-graph})\\
  $\phi$       & Relative movement labeling func. & \S~\ref{sssec:RML}\\
  $Z_{w'w}$    & Correction term & Eq.~(\ref{eq:correction term})\\\hline
 \end{tabular}
\end{table}
\subsection{Definitions}\label{sec:def}
\subsubsection{Data models}
\label{sssec:trajstr}
First, we define NCTs as follows.
\begin{defi} 
 A \emph{network-constrained trajectory} (NCT) on a directed graph $(V,E)$ is defined as a sequence of physically connected road segments, i.e., $e_1e_2\cdots e_n \; (e_i\in E)$.
\end{defi}

For example, we have $e_1=\textsf{A}$, $e_2=\textsf{B}$, $e_3=\textsf{E}$, and $e_4=\textsf{F}$ for $T_1=\textsf{ABEF}$ illustrated in Fig.~\ref{fig:main-idea}~(a).
%
To build an FM-index for a set of documents, they are usually concatenated into one long string \cite{SNT-index}.
Similarly, we define a \emph{trajectory string} that concatenates the NCTs.
\begin{defi}[Trajectory string]\label{def:trajstr}
 Let $\mathcal{T}:=\{T_k\}_{k=1}^{N}$ be a set of NCTs  to be indexed. 
 A \emph{trajectory string} is defined as
 $T:=T_1^r\bnd\hspace{1pt}T_2^r\bnd\cdots T_N^r\bnd\bsharp$,
 where $T_k^r$ is the reversal of string $T_k$, and \bnd\xspace and $\bsharp$ are special symbols that represent
 NCT boundaries and the end of the string, respectively.
\end{defi}

For the four NCTs in Fig.~\!\ref{fig:main-idea} (a), the trajectory string is
\begin{align}
 T=\underbrace{\textsf{FEBA}}_{T_1^r}\bnd
   \underbrace{\textsf{CBA}}_{T_2^r}\bnd
   \underbrace{\textsf{CB}}_{T_3^r}\bnd
   \underbrace{\textsf{DA}}_{T_4^r}\bnd\bsharp.
 \label{eq:trajstring}
\end{align}
In the later sections, we use this example for explanation.
In this paper, a string $S$ has 0-based subscripts and $|S|$ denotes its length.
$S[i]$ and $S[i,j)$ are the $i$-th element and the substring from $i$ to $j-1$, respectively.
The alphabet set is defined as $\Sigma:=E\cup\{\bnd, \bsharp\}$,
and $\sigma$ denotes its size.
To define the BWT below, we assume a lexicographical order on the road segments $E$ (any ordering can be used for our purpose).
The lexicographical order is assumed to be $\bsharp<\bnd<w \; (\forall w\in E)$.
%
%

\subsubsection{Pattern matching and BWT}
\label{sssec:bwt}
\begin{figure}[hbt]
 \begin{center}
  \vspace{-.5em}
  \includegraphics[scale=.82]{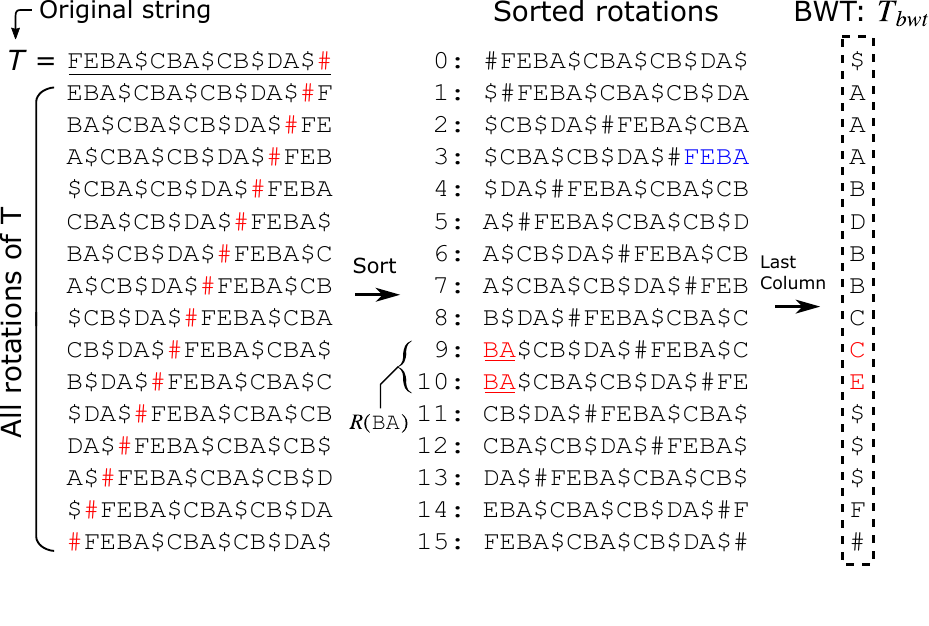}
  \vspace{-.8em}
  \caption{The BWT of $T$ is defined to be the last column of the sorted rotations of
  $T$. This example is based on the trajectory string $T$ in Eq.~(\ref{eq:trajstring}).}
  \label{fig:bwt}
  \vspace{-1.2em}
 \end{center}
\end{figure}

The \emph{Burrows--Wheeler transform} (BWT) \cite{BWT} is closely related to pattern matching and is used in FM-index.
It is a reversible transform of $T$,
defined to be the last column of the lexicographically sorted rotations of $T$ (Fig.~\ref{fig:bwt}).
For trajectory string Eq.~(\ref{eq:trajstring}), we have
\begin{align}
 T_{bwt}=\bnd\textsf{AAABDBBCCE}\bnd\bnd\bnd\textsf{F}\bsharp.
 \label{eq:bwt}
\end{align}
\hspace{-3pt}For a given pattern (string) $P$, we can define a unique range $R(P)=[sp, ep)$ for which the prefixes of the corresponding sorted rotations are equal to $P$.
We call this range the \emph{suffix range} of $P$.
For example, if $P=\textsf{BA}$, we have $R(P)=[9,11)$ (see the underlined prefixes in Lines 9 and 10 in Fig.~\ref{fig:bwt}).
Finding $R(P)$ for a given $P$ is called \emph{pattern matching}, or \emph{suffix range query}, in this paper.
The suffix range query for a trajectory string $T$ finds a suffix range of a given spatial path $P$.
It is known that the suffix range is useful in spatio-temporal query processing for NCTs (see Section \ref{sec:related-work}).
This is why we focus on this query.

In this paper, we also focus on another query, \emph{sub-path extraction query}, that recovers a sub-path of any length from an arbitrary position in BWT $T_{bwt}$.
We describe this query in Section \ref{sec:extract}.

\subsubsection{FM-index and an algorithm to find suffix ranges}
\label{sssec:fmalgo}
The FM-index \cite{FM-index} is a data structure that compresses a large string and indexes it at the same time.
Specifically, the FM-index of a string $T$ is a data structure in which the BWT of $T$ is stored in a \emph{wavelet tree}.
Suffix range queries can be processed rapidly by FM-index.
In the following, we overview how it works.
%
It is known that Algorithm~\ref{algo:fmsearch} can find the suffix range $R(P)$ for any $P$ based on $T_{bwt}$.
The \emph{rank function}, $rank_w(T_{bwt},i)$, returns the number of occurrences of a symbol $w\in\Sigma$ in a substring $T_{bwt}[0,i)$.
For example, we have $rank_{\textsf{B}}(T_{bwt},5)=1$ because
\begin{align}
 T_{bwt}=\overbrace{\bnd\textsf{AAA\underline{B}}}^{T_{bwt}[0,5)}\textsf{DBBCCE}\bnd\bnd\bnd\textsf{F}\bsharp.
 \nonumber
\end{align}
Moreover, $C[w]$ is the number of symbols in $T_{bwt}$ that are
lexicographically smaller than $w$.
For example, we have $C[\textsf{A}]=5$
and $C[\textsf{B}]=8$ by simple counting.
The range $[C[w],C[w+1])$ defines the suffix range $R(w)$:
$R(\textsf{A})=[5,8)$, for example (see that \textsf{A} appears as prefixes in $[5,8)$ in Fig.~\ref{fig:bwt}).

To understand how Algorithm~\ref{algo:fmsearch} works,
let us consider a query $P=\textsf{BA}$.
In Line 1, we have $w=\textsf{A}$, $sp=5$, and $ep=8$.
Consider the first (and last) iteration with $i=2$.
We have $sp=C[\textsf{B}]+1=9$ and $ep=C[\textsf{B}]+3=11$
because $\textit{rank}_{\textsf{B}}(T_{bwt},sp)=1$ and
$\textit{rank}_{\textsf{B}}(T_{bwt},ep)=3$ by definition.
Therefore $[sp,ep)=[9,11)$ is returned at Line 7,
which is equivalent to $R(\textsf{BA})$ given in Fig.~\ref{fig:bwt}.

\begin{figure}[tb]
 \begin{center}
  \vspace{-.5em}
  \includegraphics[width=7.4cm,height=3.7cm]{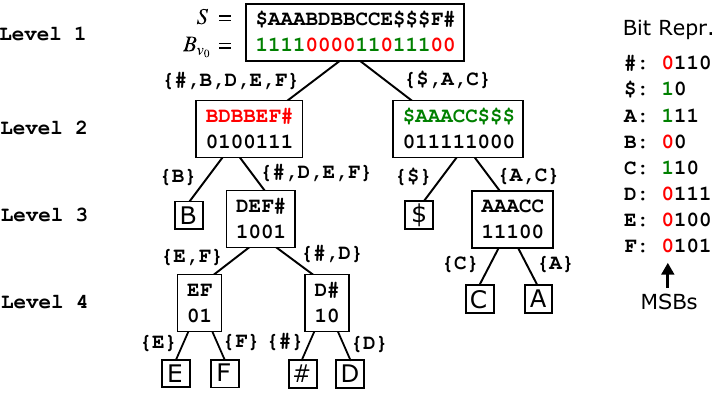}
  \caption{Wavelet tree: 
  \textnormal{a bit representation of each symbol
  in a string $S$ is stored in a binary tree
  (this example is the HWT of the BWT of the trajectory string Eq.~(\ref{eq:bwt})).
  Note that only bit vectors are
  stored in each node.
  }}\label{fig:wt}
 \vspace{-1em}
 \end{center}
\end{figure}

We can say that fast calculation of $\textit{rank}_w$ enables
the fast execution of Algorithm \ref{algo:fmsearch} 
because all the operations except for
$\textit{rank}_w(T_{bwt},i)$ are merely either substitutions or summations.
However, na\"{i}ve calculation of $\textit{rank}_w$ with cumulative counting incurs an unacceptable $O(|T_{bwt}|)$ time.
\begin{algorithm}[tb]
   \caption{\small Finding the suffix range $R(P)=[sp, ep)$
   for a given query $P$ of length $m$
   based on $T_{bwt}$ (\textit{SearchFM})}
   \label{algo:fmsearch}
\DontPrintSemicolon 
\KwIn{BWT string of length $n$: $T_{bwt}$,\\
 \hspace{2.7em} Query string of length $m$: $P$}
\KwOut{Range of $T_{bwt}$ that matches to $P$}
   $w\leftarrow P[m-1]$; $sp\leftarrow C[w]$;
   $ep\leftarrow C[w+1]$\;
   \For{$i\leftarrow 2$ to $m$}{
   $w\leftarrow P[m-i]$\;
   $sp\leftarrow C[w]+\textit{rank}_w(T_{bwt}, sp)$\;
   $ep\leftarrow C[w]+\textit{rank}_w(T_{bwt}, ep)$\;
   \lIf{$sp\ge ep$}{
   \Return \texttt{NotFound}
   }
   }
 \Return $[sp, ep)$
\end{algorithm}
\mypara{Wavelet tree}\label{sec:WTintro}
A \textit{wavelet tree} \cite{WT} storing $T_{bwt}$ enables
fast calculation of $\textit{rank}_w(T_{bwt},i)$;
its time complexity does not depend on the data size $|T_{bwt}|$.
Figure~\ref{fig:wt} illustrates a wavelet tree for the string
$S=T_{bwt}$ in Eq.~(\ref{eq:bwt}).
The bit representation of each symbol is predefined
(e.g., Huffman coding based on the frequency of each symbol
in $T_{bwt}$).
Each node $v$ in the tree stores a bit vector $B_v$.
For the root node $v_0$, $B_{v_0}$ stores the most significant
bit (MSB) of each symbol in $S$.
At the second level, the symbols are divided into two parts
based on the bit value at the first level, while keeping the ordering.
Each bit vector stores the second MSB.
Repeating such partitioning recursively, we obtain the wavelet tree.
In fact, $B_v$ is stored in a \emph{succinct dictionary}
\cite{UncompressedBitVector,RRR}, which is
a bit vector that supports a bit-wise rank
(i.e., $\textit{rank}_0(B_v,j)$ and $\textit{rank}_1(B_v,j)$)
in $O(1)$ time.

There are several types of wavelet tree with different compression
characteristics that are determined by tree shape and the type of
succinct dictionary \cite{WTAll}.
In \proposedsf, we use a \textit{Huffman-shaped wavelet tree}
(HWT) \cite{HuffmanWT}, whose tree shape
is that of the Huffman tree of $S$.
It is known that an HWT can compress a string $S$ of length $n$ to
at most $n(1+H_0(S))+o(n)$ bits.
Here, $H_0(S)$ is the \textit{0th order empirical entropy}
\cite{EmpiricalEntropy},
\begin{align}
 H_0(S)=\sum_{w\in\Sigma}\frac{n_w}{n}\lg\frac{n}{n_w},
 \label{eq:H0}
\end{align}
where $n_w$ is the number of occurrences of $w$ in $S$.

To calculate $\textit{rank}_w(S,j)$, the wavelet tree calculates the bit-wise rank value
at each node $v_0, v_1, \cdots, v_k$ between the root and the leaf corresponding to the bit representation $w=b_0b_1\cdots b_k$ (see \cite{WT} for details).
This indicates that bit-wise rank operations required to obtain $\textit{rank}_w(S,j)$ is equal to $k$ (i.e., the length of the bit representation of $w$).
This fact leads to the following result \cite{WTAll}.
\begin{theo}[Rank on HWT]\label{theo:rank-huff-wt}
If $\textit{rank}_w(S,j)$ is executed on uniformly random $w$ over $S[0,n)$, it runs in $O(1+H_0(S))$ time on average.
\end{theo}

This result implies that a string with small entropy $H_0(S)$
achieves not only small size but also fast rank operation,
which plays an important role in our theoretical analysis.

\subsection{Compressed variants of FM-index}\label{sec:compboost}
Let us consider a sub-path of length 3 in a real NCT dataset:
$e_{t-2}\,e_{t-1}\,e_t$.
It is unlikely that two right turns occur in a row
because most vehicles go toward their destinations.

\begin{figure}[t]
 \begin{center}
  \vspace{-.5em}
  \includegraphics[scale=.95]{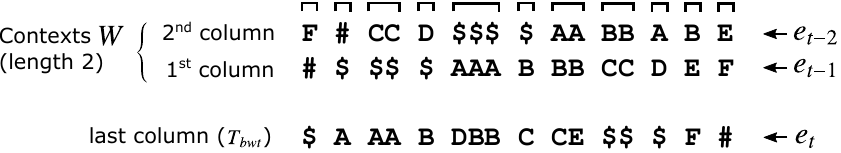}
  \caption{Compression boosting of FM-index:
  \textnormal{$T_{bwt}$ is divided into \emph{contexts} and each
  partition is compressed separately.}}
  \label{fig:compboost}
 \end{center}
 \vspace{-2em}
\end{figure}

Considering such high-order correlations among symbols, we can boost the compression.
As noted before,
the prefix \textsf{BA}\,$\in\!\!\Sigma^2$ appears in $[9,11)$
(Fig.~\ref{fig:bwt}).
The other prefixes $W\!\in\!\Sigma^2$ have their corresponding ranges.
Let us divide $T_{bwt}$ based on such prefixes $W$
(called \emph{contexts} of length two)
as shown in Fig.~\ref{fig:compboost}.
These context blocks represent the next segment $e_t$
given the context $W=e_{t-1}\,e_{t-2}$.
We have a chance of compression because, as discussed above, the frequency of symbols in each context is biased.
\mypara{Compression boosting (CB)}
The above idea can be generalized to any length of context.
Let us divide $T_{bwt}$ into $l$ blocks
of context $W\in\Sigma^k$ of length $k$:
$T_{bwt}=L_1L_2\cdots L_l$ ($l\le\sigma^k$).
Storing each $L_j$ in a $0$-th order entropy compressor such as an HWT,
we can compress $T_{bwt}$ to $nH_k(T)+o(n)$.
Here, $H_k$ is $k$-th order empirical
entropy
\cite{EmpiricalEntropy}:
\begin{align}
 H_k(T):=\sum_{W\in\Sigma^k}\frac{n_W}{n}H_0(T_W),
 \label{eq:Hk}
\end{align}
where $T_W$ is the concatenation of all symbols in $T$
that precede 
the context $W$.
To support a fast rank operation on those divided blocks,
we need to precompute and store the rank results at each location of
$l$ blocks for all $w\in\Sigma$.

Taking larger $k$ seems to be desirable because $H_{k}(T)\ge H_{k+1}(T)$ for all $k\ge0$ \cite{EmpiricalEntropy}.
However, partitioning into many blocks leads to the following problems in practice:
\begin{enumerate}
 \setlength{\itemsep}{-0cm}
 \item[\textit{P1)}] Blocks of variable length lead to inefficient
 	    random access to $T_{bwt}$.
 \item[\textit{P2)}] Index size increases because of the overhead of block-wise
	    storage (e.g., pointers in Huffman trees).
 \item[\textit{P3)}] 
	    We have to save $l\sigma$ integers for the rank results.
	    This is unrealistic for huge $\sigma$ even if $k=1$ ($l=\sigma$).
\end{enumerate}

\mypara{Variants of CB}
There are some CB variants that avoid the above problems.
Fixed-block compression boosting \cite{FCBC} adopts blocks of a
fixed size. Although this solves P1 (and P2 partially),
problem P3 remains for huge $\sigma$.
\emph{Implicit compression boosting} (ICB) \cite{ImplicitCompBoost} avoids such explicit block partitioning by using
a compressed succinct dictionary called an \emph{RRR} \cite{RRR} in the wavelet tree of $T_{bwt}$.
This implicit partition solves P1 and P3. 
Brisaboa et al.~\cite{CTR} employed ICB to index a trajectory string of NCTs.
Specifically, they employed ICB with a \emph{wavelet matrix} \cite{WM}, which is an efficient alternative to a wavelet tree.
We call this structure \textsf{ICB-WM} in this paper (similarly, we refer to ICB with an HWT as \textsf{ICB-Huff}).
As discussed in our theoretical analysis, ICBs still suffer from large overheads when applied to a string with large alphabet, such as a trajectory string of NCTs.

\section{Proposed data structure}
\label{sec:proposed}
\subsection{Overview}\label{sec:prop overview}
\begin{figure}[t]
 \vspace{-.0em}
 \begin{center}
  \includegraphics[scale=0.65]{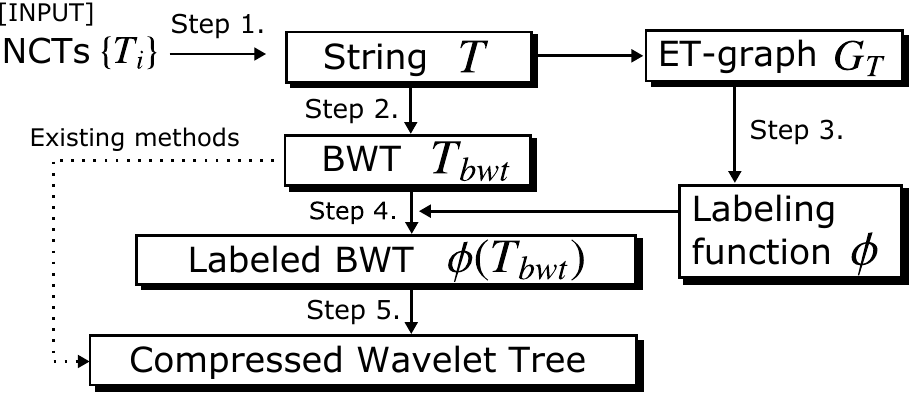}
  \caption{Overview of \proposedsf (index construction). Note that the existing ICBs do not have the labeling step.}
  \label{fig:method_overview}
  \vspace{-1em}
 \end{center}
\end{figure}

For NCTs, the alphabet size $\sigma$ can be millions because it is the number of road segments in a road network.
As discussed in the previous section, this makes the compression of trajectory strings inefficient, because the redundant bits in wavelet trees increase as $\sigma$ increases.
To avoid this, we convert trajectory strings into strings with a small alphabet via \emph{relative movement labeling} (RML), which is based on the sparsity of road networks.
%
Figure~\ref{fig:method_overview} and the following give an overview of how to construct the proposed data structure, \proposedsf.
\begin{enumerate}
 \setlength{\itemsep}{0.03cm}
 \item Convert a set of NCTs into a trajectory string $T$.
 \item Calculate the BWT of $T$ and obtain $T_{bwt}$.
 \item Construct an \emph{ET-graph} $G_T$ and a \textit{relative movement labeling} (RML) function $\phi$ based on $T$ (Section~\ref{sec:traj encoding})
 \item Label $T_{bwt}$ based on the RML function $\phi$ and obtain the \textit{labeled BWT} $\phi(T_{bwt})$ (Section~\ref{sec:trajwt}).
 \item Store $\phi(T_{bwt})$ in an HWT with RRR and obtain the proposed index structure (Section~\ref{sec:trajwt}).
\end{enumerate}

As steps 1 and 2 are straightforward,
we describe the details of steps 3--5 in the following sections.
We emphasize that the NCTs are labeled \textit{after} the BWT (step 4), otherwise we would be unable to implement the suffix range query.
Due to this labeling step, we need to develop an algorithm that differs from Algorithm \ref{algo:fmsearch}.
Such an algorithm is described in Section \ref{sec:search}.
The theoretical consequences of \proposedsf are described in Section~\ref{sec:theory}.
Here, we focus on the index structure.
%

Note that \proposedsf basically deals with static data.
We can treat growing data by periodic reconstruction or by constructing an index for new data at certain time intervals.
%
\subsection{Relative movement labeling (RML)}\label{sec:traj encoding}
The RML converts trajectory strings into strings with small alphabet based on the following fact: \emph{NCTs can only move between physically connected road segments}.
First, we describe its idea based on the example in Fig.~\ref{fig:main-idea} (a).
If a vehicle is on a road segment $w'=\textsf{A}$, the next segment $w$ has to be $\textsf{B}$ or $\textsf{D}$.
Hence, we label them \texttt{1} and \texttt{2}, respectively.
Generally, if there are $k$ connected road segments from a certain segment, we can label them with \texttt{1}, $\cdots$, \texttt{k}.
The sequences converted with this \emph{relative movement labeling} (RML) are expected to have small alphabet because $k$ is smaller than the maximum out-degree of the road network.
%
%
%
%
To define RML formally, let us define an \textit{empirical transition graph} (ET-graph).

\begin{defi}[ET-graph]\label{def:ET-graph}
 Let $T$ be a string defined on an alphabet $\Sigma$.
 An \emph{ET-graph} $G_T$ of $T$ is a directed graph that satisfies:
 \emph{1)} the vertex set is $\Sigma$;
 \emph{2)} a directed edge $(w', w)\in\Sigma\times\Sigma$ exists iff there exists a substring $ww'$ in $T$.
 The edge set is denoted by $E_T$.
\end{defi}

\begin{figure}[tb]
 \begin{center}
  \subfigure[ET-graph and RML]{
    \includegraphics[scale=1.1,trim=-.2em 0 -.2em 0]{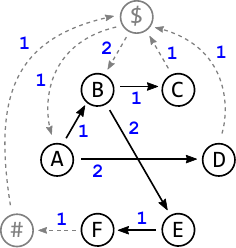}
  }
  \hspace{4pt}
  \subfigure[Labeling $T_{bwt}$ with RML]{
  \includegraphics[scale=.95,trim=0 -5pt 0 0]{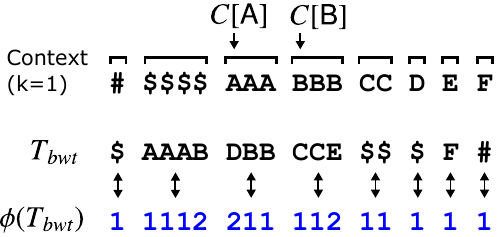}
  }
  \caption{(a) ET-graph of our example Eq.~(\ref{eq:trajstring}):
  \textnormal{each node represents a
  road segment, and an edge exists if the corresponding transition
  occurs in $T$. The integer on each edge is the corresponding
  label.} (b) Relative movement labeling: \textnormal{$\phi(T_{bwt})$
  produces a string with a lower entropy than that of $T_{bwt}$.}}
  \label{fig:edge code}
  \vspace{-1em}
 \end{center}
\end{figure}

In other words, an edge exists iff a direct transition between $w'$ and $w$ exists in $T$.
The ET-graph is a sparse graph because it has a similar topology to the original road network.
Figure~\ref{fig:edge code}~(a) illustrates the ET-graph of
the trajectory string $T$ given in Eq.~(\ref{eq:trajstring}).
Note that ET-graphs include the special symbols \bnd\xspace and \bsharp.

\mypara{Definition of RML}\label{sssec:RML}
The RML can be defined as an integer assigned on each edge of ET-graph (see Fig.~\ref{fig:edge code}~(a)).
For example, the transition $\textsf{A}\to\textsf{B}$ is labeled \texttt{1}.
The transition $\textsf{A}\to\textsf{D}$ must have the different label, otherwise we cannot distinguish them.
For transition $w'\to w$, we denote such a labeling function by $\phi(w|w')$.
For example, we have $\phi(\textsf{B}|\textsf{A})=\texttt{1}$ and $\phi(\textsf{D}|\textsf{A})=\texttt{2}$.
To make the labeling \emph{distinct} based on the previous symbol $w'$, the RML function $\phi$ must satisfy the following requirement.
\begin{itemize}
 \item \textit{Requirement}:
\textit{The RML function $\phi(\cdot|w')$ must be a one-to-one map for any $w'$.}
\end{itemize}

Now, we discuss how to construct the RML function $\phi$ that satisfies the requirement above.
Let us consider the \textit{out-vertex set} of $w'$, defined as $N_{out}(w')=\{w|(w',w)\in E_T\}$,
that determines the set of vertexes directly accessible from $w'$.
Based on the ET-graph and out-vertex set, we define $\phi(\cdot|w')$ as follows.
Given $w'$, assign a different small integer $c_{ww'}$ to each $w\in N_{out}(w')$ and define $\phi(w|w'):=c_{ww'}$.
It is clear that $\phi(\cdot|w')$ is a one-to-one map.
If $w\notin N_{out}(w')$, we cannot define $\phi(w|w')$.
However, this is not a problem because $w\notin N_{out}(w')$ indicates that the string $ww'$ is not found in $T$,
which tells us the result of pattern matching is null.
This point is important for our search algorithm.
\mypara{Finding an optimal RML}


The RML $\phi$ described above does not define a unique labeling function because we have not yet specified
a concrete way to assign the small integers $c_{ww'}$.
Here, we propose a strategy based on a bigram count $n_{ww'}$ (i.e., the frequency of $ww'$ in $T$).
The elements in $N_{out}(w')$ are sorted in descending order of bigrams $n_{ww'}$.
The vertex $w$ with the largest bigram count is given the smallest label, \texttt{1}.
The second-most frequent vertex is labeled \texttt{2}, the third-most frequent vertex is labeled \texttt{3}, and so on.
The labels shown in Fig.~\ref{fig:edge code} (a) are determined in this way.
For example, since we have $n_{\textsf{BA}}\!>\!n_{\textsf{DA}}$ ($n_{\textsf{BA}}\!=\!2$ and $n_{\textsf{DA}}\!=\!1$), the edge from \textsf{A} to \textsf{B} has the smallest label \texttt{1}: $\phi(\textsf{B}|\textsf{A})=\texttt{1}$.
Applying a labeling scheme shown in the next section, this labeling strategy generates a low-entropy sequence $\phi(T_{bwt})$ as shown in Fig.~\ref{fig:edge code} (b), because the distribution of the resulting symbols is biased toward smaller integers (i.e., \texttt{1} is the largest fraction).
For this example, we have $H_0(T_{bwt})=2.8$ and $H_0(\phi(T_{bwt}))=0.7$ (unit: bits).

One might wonder whether there exists a better labeling strategy.
We prove, however, the optimality of the labeling that leads to strong conclusions:
\emph{our RML achieves the smallest size and the fastest search}.
See Section~\ref{sec:optimality} for details.

\subsection{Data structure}\label{sec:trajwt}
Here, we describe how to obtain $\phi(T_{bwt})$ and the final index (steps 4 and 5 in Section \ref{sec:prop overview}).
\mypara{Labeling BWT (step 4)}
Based on the RML function $\phi$ obtained in the previous section, the BWT $T_{bwt}$ is converted to $\phi(T_{bwt})$ in the following manner.
For example, let us focus on the third block of $T_{bwt}$, \textsf{DBB}, in Fig.~\ref{fig:edge code} (b).
This block corresponds to the context of \textsf{A}, which indicates that the previous symbol of these \textsf{DBB} is \textsf{A}.
Hence, \textsf{DBB} is labeled as \texttt{211} because $\phi(\textsf{B}|\textsf{A})=\texttt{1}$ and $\phi(\textsf{D}|\textsf{A})=\texttt{2}$ in Fig.~\ref{fig:edge code} (a).
All the other blocks also can be labeled in the same manner.

\begin{figure}[bt]
 \begin{center}
  \subfigure[$T_{bwt}$]{
  \includegraphics[scale=1.05]{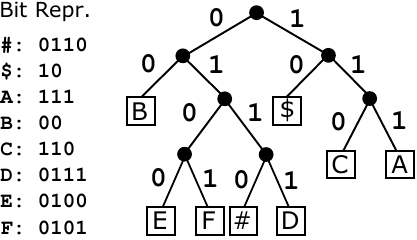}
  }
  \hspace{1em}
  \subfigure[$\phi(T_{bwt})$: \proposedsf]{
  \includegraphics[scale=1.05,trim=0 -1.8em 0 0]{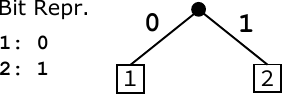}
  }
  \caption{
  Huffman trees of $T_{bwt}$ and $\phi(T_{bwt})$:
  \textnormal{each leaf corresponds to a symbol.
  The tree generated by \proposedsf (b) is simpler than that from the existing technique (a).
  }
  }
  \label{fig:hufftree}
  \vspace{-1em}
 \end{center}
\end{figure}

\mypara{Storing to a compressed wavelet tree (step 5)}
In this step, we store the labeled BWT $\phi(T_{bwt})$ to an HWT.
For bit vectors in an HWT, we adopt a practical version of the compressed succinct dictionary called RRR \cite{PracticalRRR}.
This is a straightforward step.
Figure~\ref{fig:hufftree} depicts the comparison of Huffman trees of $T_{bwt}$ and $\phi(T_{bwt})$ for the example in Figure \ref{fig:edge code} (b).
The Huffman tree of $\phi(T_{bwt})$ is obviously simpler than that of $T_{bwt}$.
Because these tree shapes are the same as those of HWTs, this simplification explains intuitively why \proposedsf is small and fast.
For more details, see Section~\ref{sec:theory}.

An RRR bit vector has one parameter $b$,
that controls the size of the internal blocks.
For larger $b$, we obtain
better compression but slower search ($rank$ calculation) in general,
and \textit{vice versa}.
This $b$ is the only parameter in \proposedsf.
However, in Section \ref{sec:experiment},
we show that this parameter has only a small influence on the index size
and the search time.
%
\mypara{Storing ET-graph}
We use an adjacency list to represent the ET-graph $G_T$.
The value $\phi(w|w')$ is assigned
to the edge $(w',w)\in E_T$.
Thus we can obtain $\phi(w|w')$
in $O(\delta)$ time by a linear search over $N_{out}(w')$.
We also assign $C[w]$ to each vertex $w$ in $G_T$.
\textit{Correction terms} $Z_{w'w}$, which are introduced
in Section~\ref{sec:pseudorank}, are also attached to
$(w',w)\in E_T$.
Note that, since $G_T$ is sparse, the space needed to store $G_T$ is negligible when $|T|$ gets large.

\section{Proposed query processing algorithms}
\label{sec:search}
Here, we describe another key concept of this paper, \emph{PseudoRank}, then show algorithms for two types of queries, suffix range queries and sub-path extraction queries.
\subsection{PseudoRank}
\label{sec:pseudorank}
As mentioned in Section~\ref{sssec:fmalgo}, fast calculation of $\textit{rank}_w(T_{bwt},j)$ is
needed for Algorithm~\ref{algo:fmsearch}.
The original FM-index stores $T_{bwt}$ in a wavelet tree to calculate ranks quickly.
In our case, however, we do not have the original $T_{bwt}$ but only have the labeled $\phi(T_{bwt})$.
Can we obtain the rank values for the original BWT by using only the labeled BWT?
Seemingly, this is difficult because different symbols are mapped to the same label
(e.g., both \textsf{A} and \textsf{C} are converted to \texttt{1}
as illustrated in Fig.~\ref{fig:edge code} (b)).

%

\begin{figure}[tb]
 \begin{center}
  \vspace{-.5em}
  \includegraphics[scale=1.1]{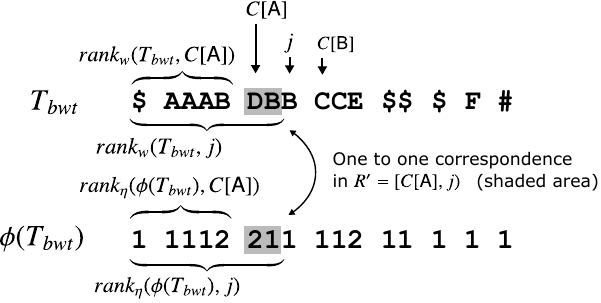}
 \caption{Basis of the balancing equation
  (Eq.~(\ref{eq:local-balance})) for \emph{PseudoRank}}
 \label{fig:pseudo_rank}
 \end{center}
 \vspace{-1em}
\end{figure}

The key idea in \proposedsf is to simulate the rank operation over $T_{bwt}$.
Figure~\ref{fig:pseudo_rank} illustrates this idea.
Let us consider the range
$R(\textsf{A})=[C[\textsf{A}],C[\textsf{B}])$
and $j\in R(\textsf{A})$.
Because the substring $T_{bwt}[C[\textsf{A}],C[\textsf{B}])=\textsf{DBB}$
is labeled as $\texttt{211}$ by using the one-to-one map
$\phi(\cdot|\textsf{A})$
as described in Section~\ref{sec:trajwt},
the following two counts are equivalent for $\forall j\in R(\textsf{A})$:
\begin{itemize}
 \item the number of occurrences of \textsf{D}
       within the range $R':=[C[\textsf{A}],j)$
       in $T_{bwt}$ (the shaded region
       in Fig.~\ref{fig:pseudo_rank}), and
 \item the number of occurrences of \texttt{2} within $R'$ in $\phi(T_{bwt})$.
\end{itemize}
This balancing relationship holds in general.
Let us consider a context $w'$.
For all $j$ such that $C[w']\le j\le C[w'\!+1]$,
let us consider a range $R':=[C[w'],j)$. 
For a symbol $w\in N_{out}(w')$,
the number of occurrences $w$ within $R'$ in $T_{bwt}$ and
that of the label $\eta:=\phi(w|w')$ within $R'$ in $\phi(T_{bwt})$ 
are the same because of the one-to-one requirement for $\phi(\cdot|w')$.
This leads to the following balancing equation:
\begin{align}
 \nonumber
 &
 \textit{rank}_w(T_{bwt},j)-\textit{rank}_w(T_{bwt},C[w'])\\
 &
 \quad=
 \textit{rank}_{\eta}(\phi(T_{bwt}),j)
 -\textit{rank}_{\eta}(\phi(T_{bwt}),C[w']).
 \label{eq:local-balance}
\end{align}
Rearranging this equation, we have the following theorem,
which allows us to simulate the rank operation.
\begin{theo}[Pseudo-rank]
 \label{thm:pseudorank}
 If $w\in N_{out}(w')$ and\\ $C[w']\le j\le C[w'\!+1]$, then we have
 \begin{align}
 &\!\!\textit{rank}_w(T_{bwt},j)=\textit{rank}_{\eta}(\phi(T_{bwt}),j)
 -Z_{w'w},\label{eq:pseudo rank}
 \\
 \nonumber
 & \!\text{where}\quad \eta:=\phi(w|w') \;\;\text{and}\\
 & \!Z_{w'w}
 := \textit{rank}_{\eta}(\phi(T_{bwt}),C[w']) 
  - \textit{rank}_w(T_{bwt},C[w']). \label{eq:correction term}
 \end{align}
\end{theo}

We emphasize that the correction term $Z_{w'w}$ does not depend on $j$,
implying that the number of correction terms needed
is equal to $|E_T|$.
Importantly, this property allows us to
precompute and store the correction terms (as noted in Section
\ref{sec:trajwt}, they are attached to each edge $(w',w)\in E_T$).

This theorem produces Algorithm~\ref{algo:pseudorank},
which calculates the rank values 
using only $\phi(T_{bwt})$.
We also emphasize that \textit{PseudoRank}
does not allow us to calculate rank values for all pairs of $(w,j)$.
However, this limitation is not a problem for our search algorithm,
as shown in the next subsection.
  \begin{algorithm}[hb]
   \caption{\small Pseudo calculation of $rank_w(T_{bwt},j)$ by using only
    $\phi(T_{bwt})$ 
   (\textit{PseudoRank}$(\phi(T_{bwt}),j,w,w',Z_{w'w})$)}
   \label{algo:pseudorank}
   \DontPrintSemicolon 
   \KwIn{Labeled BWT string of length $n$: $\phi(T_{bwt})$,\\
   \hspace{3.2em} Location of rank $j$,
   \hspace{.5em} Correction term $Z_{w'w}$,\\
   \hspace{3.2em} Target symbol $w$,
   \hspace{.5em} Previous symbol $w'$}
   \KwOut{The value of $rank_w(T_{bwt},j)$}
   \If{$w\in N_{out}(w')$ and $C[w']\le j\le C[w'\!+1]$}{
   $\eta\leftarrow\phi(w|w')$\tcp*[r]{RML}
   \Return $\textit{rank}_{\eta}(\phi(T_{bwt}),j)-Z_{w'w}$
   }
   \Return \texttt{NotFound}
  \end{algorithm}

\subsection{Suffix range query with \proposed}
\label{sec:proposed algo}
With the \textit{PseudoRank},
we can simulate $\textit{rank}_w(T_{bwt},j)$
using only the wavelet tree of $\phi(T_{bwt})$
and the correction term $Z_{w'w}$ (Eq.~(\ref{eq:correction term})).
Replacing the rank operations in Algorithm~\ref{algo:fmsearch}
with \textit{PseudoRank}, we obtain our search algorithm
(Algorithm~\ref{algo:encoded fmsearch}), whose
correctness is shown below.
\mypara{Correctness of the algorithm}
To guarantee that Algorithm~\ref{algo:encoded fmsearch}
is equivalent to Algorithm~\ref{algo:fmsearch},
we have to check the following two conditions
on \textit{PseudoRank} (Theorem~\ref{thm:pseudorank})
are satisfied immediately before Line 7:
(c1) $w\in N_{out}(w')$;
(c2) $C[w']\le sp\le C[w'\!+1]$ and $C[w']\le ep\le C[w'\!+1]$.

As noted previously, no substring $ww'$ appears in $T$ if
$w\notin N_{out}(w')$;
hence, \texttt{NotFound} is returned if
$w\notin N_{out}(w')$ at Line 6.
Therefore, (c1) $w\in N_{out}(w')$ holds immediately before Line 7.
For (c2),
before Line 7, $sp$ satisfies
\begin{align}
 sp=C[w']+\textit{rank}_{w'}(T_{bwt},sp'),
 \label{eq:st}
\end{align}
where $sp'$ is the previous value.
By the \textit{rank} definition,
\begin{align}
 0\le \textit{rank}_{w'}(T_{bwt},j)\le C[w'\!+1]-C[w'],\quad 0\le\forall j<|T|
 \nonumber
\end{align}
holds, where $C[w'\!+1]-C[w']$ means the number of occurrences of $w'$ in $T$.
Combining this inequality with Eq.~(\ref{eq:st}),
we obtain $C[w']\le sp\le C[w'\!+1]$.
We can prove the condition for $ep$ in a similar manner.

\begin{algorithm}[tb]
   \caption{\small Finding the suffix range $[sp, ep)$
   for a given query $P$ of length $m$
   based on $\phi(T_{bwt})$ (\textit{LabeledSearchFM})}
   \label{algo:encoded fmsearch}
\DontPrintSemicolon 
\KwIn{Labeled BWT string of length $n$: $\phi(T_{bwt})$,\\
 \hspace{3.2em} Query of length $m$: $P[0,m)$,\\
 \hspace{3.2em} Correction terms: $\{Z_{w'w}\}$}
\KwOut{Range of $T_{bwt}$ that matches to $P$}
   $w\leftarrow P[m-1]$;
   $sp\leftarrow C[w]$;
   $ep\leftarrow C[w+1]$\;
 \For{$i\leftarrow 2$ \textsf{\upshape to} $m$}{
 $w'\leftarrow w$\tcp*[r]{Save previous symbol}
 $w\leftarrow P[m-i]$\;
 \If{$w\notin N_{out}(w')$}{\Return \texttt{NotFound}}
 $sp\leftarrow\!C[w]\!+\!\textit{PseudoRank}(\phi(T_{bwt}), sp, w, w', Z_{w'w})$\;
 $ep\leftarrow\!C[w]\!+\!\textit{PseudoRank}(\phi(T_{bwt}), ep, w, w', Z_{w'w})$\;
 \If{$sp\ge ep$}{
 \Return \texttt{NotFound}
 }
 }
 \Return $[sp, ep)$
\end{algorithm}
\subsection{Extracting a sub-path with \proposed}
\label{sec:extract}
%
Here, we describe another important query, \emph{sub-path extraction query}.
For example, let us focus on the \emph{third} sorted rotation in Fig.~\ref{fig:bwt}.
Its suffix of length four is \texttt{FEBA} (colored in blue).
This corresponds to the example NCT $T_1^r$ in Eq.~(\ref{eq:trajstring}).
In this way, the sub-path extraction queries recover a sub-path of length $l$ from an arbitrary position $j$ in BWT $T_{bwt}$ ($j=3$ for the example above).
This query is useful if we need to obtain certain NCTs stored in BWT string, or we need to recover the entire trajectory string.
Formally, $\textit{extract}(j, l)$ returns $T[i-l,i)$ where $i=SA[j]$ ($SA$ is the suffix array of $T$). 
The subscript $j$ is often referred to as \emph{inverse suffix array} ($j=ISA[i]$).

Algorithm~\ref{algo:recover} shows how to obtain $\textit{extract}(j, l)$ using only $\phi(T_{bwt})$ and the ET-graph.
This is obtained by mimicking \emph{LF-mapping} \cite{FM-index} with PseudoRank.
Line~1 performs a binary search to find the last character $T[i]=w'$ such that $C[w']\le j<C[w'\!+1]$.
Line~4 first accesses the $j$-th character of $\phi(T_{bwt})$
(i.e., the labeled $T_{bwt}[j]$),
then decodes the $T_{bwt}[j]=T[i-k-1]=w$ using the ET-graph.
Line~5 is similar to Line~7 in Algorithm~\ref{algo:encoded fmsearch},
which jumps to the next position on $T_{bwt}$ (LF-mapping simulated by PseudoRank).
\begin{algorithm}[tb]
   \caption{\small Extracting a sub-path $T[i-l+1]\cdots T[i-1]T[i]$
   for given $j=ISA[i]$ and $l>0$ (\emph{extract})}
   \label{algo:recover}
\DontPrintSemicolon 
\KwIn{Labeled BWT: $\phi(T_{bwt})$,
 Position on $T_{bwt}$: $j$,\\
 Extraction length: $l$,
 Correction terms: $\{Z_{w'w}\}$}
 \KwOut{A substring $S:=T[i-l+1]\cdots T[i-1]T[i]$} 
 $w'\leftarrow \textit{BinarySearch}(j, \{C[w']\})$\tcp*[r]{T[i]}
 \For{$k\leftarrow 1$ \textsf{\upshape to} $l$}{
 $\eta\leftarrow\phi(T_{bwt})[j]$;
 $w\leftarrow\textit{decode}(\eta|w')$;
 $S[l-k]\leftarrow w$\;
 $j\leftarrow C[w]+\textit{PseudoRank}(\phi(T_{bwt}), j, w, w', Z_{w'w})$\;
 $w'\leftarrow w$\tcp*[r]{Save previous symbol}
 }
 \Return $S$
\end{algorithm}

%
%
%
%

\section{Theoretical analysis}
\label{sec:theory}
In this section, we explain theoretically why \proposedsf is compact and fast.
We first show the optimality of our proposed RML, that is, the labeled BWT $\phi(T_{bwt})$ achieves the smallest entropy.
Then, we explain that such a small entropy contributes high compressibility and fast query processing.
We also show that RML is better than other labeling method called \LandC, recently proposed in \cite{COMPRESS}.
\subsection{Optimality of RML}
\label{sec:optimality}
The 0th order empirical entropy $H_0$ given in Eq.~(\ref{eq:H0}) plays important roles in our analysis.
First, we show the labeling strategy based on bigram counts $n_{ww'}$ proposed in Section~\ref{sec:traj encoding} achieves the minimum value of $H_0$ among all possible labelings.

\begin{theo}[Optimality]
 \label{theo:optimality}
 Let $\phi^*$ be the RML based on the bigram ordering strategy 
 and $\phi$ be any possible RML that satisfies the requirement in Section~\ref{sec:traj encoding}.
 Then, we have
 \begin{align}
  H_0(\phi^*(T_{bwt}))\le H_0(\phi(T_{bwt})).
 \end{align}
\end{theo}
\begin{IEEEproof}
 See Appendix \ref{sec:optimality-proof}.
\end{IEEEproof}

As a special case of this theorem, we obtain an \emph{unlabeled case} result, i.e., $H_0(\phi^*(T_{bwt}))\le H_0(T_{bwt})$, by putting as $\phi=id$ (identity labeling).
Importantly, we see that 
\begin{align}
 H_0(\phi^*(T_{bwt}))\ll H_0(T_{bwt})
 \label{eq:entropy-ineq}
\end{align}
holds for real NCT datasets in our experiments (Table \ref{table:dataset-stat}).
\subsection{Compressed size}
\label{sec:theory:q1}
\mypara{Evaluating space overheads}
The data structure of \proposedsf consists of two parts: the labeled BWT $\phi(T_{bwt})$ and the ET-graph $G_T$.
As noted in Section~\ref{sec:trajwt}, the size of $G_T$ is negligible when $|T|$ is large.
Here, we compare the sizes of $T_{bwt}$ and $\phi(T_{bwt})$ stored in HWTs with RRR.
Note that these corresponds \textsf{ICB-Huff} and \proposedsf, respectively.
%
The main advantage of \proposedsf comes from the lower space overhead due to RRR, as explained below.
For a given bit vector $B$, the practical RRR with the parameter $b$ uses at most
\begin{align}
 |B|H_0(B)+|B|\cdot h(b)
 \label{eq:rrr}
\end{align}
bits \footnote{\small In fact, there exist non-dominant terms that are not included in
this equation. See \cite{PracticalRRR} for details.} where $h(b)=\frac{\lg(b+1)}b$ \cite{PracticalRRR}.
We call the second term the \emph{RRR-overhead}.
For $b=63$, we have an overhead of $h(b)=(\lg 64)/63 \simeq 0.095$ bits per bit.

For a given string $S$, the average code length with Huffman coding is at most $(1+H_0(S))$ bits.
Hence, the total length of bit vectors in the HWT is
$\sum_v|B_v|\simeq |S|(1+H_0(S))$ (Section~\ref{sec:def}).
Summing the RRR-overheads over all internal nodes $v$ in the HWT,
we obtain total bits of the overhead:
\begin{align}
 \sum_v|B_v|\cdot h(b) \simeq |S|(1+H_0(S))\cdot h(b).
 \label{eq:proposed_overhead}
\end{align}
The right-hand side implies that the
RRR-overhead of a sequence $S$ is small if
its entropy $H_0(S)$ is small.
Therefore, Eq.~(\ref{eq:entropy-ineq}) indicates that the space overhead for \proposedsf is much smaller than that for \textsf{ICB-Huff}.
\mypara{High-order compression}
Here, we analyze the remaining first (and dominant) term in Eq.~(\ref{eq:rrr}).
Summing this term over all internal nodes $v$ in the HWT, we find that the total bits needed for this term achieves $k$-th order entropy Eq.~(\ref{eq:Hk}) for all $k>0$.
This property implies that our method guarantees a high compressibility in information theoretic sense.
Note that this kind of entropic bound has not been guaranteed by the existing shortest-path based NCT compressors.
%
\begin{theo}
 \label{theo:proposedCB}
 For all $k>0$, the total bits required to store $\phi(T_{bwt})$ in an HWT with RRR,
 apart from the overhead Eq.\!\!~(\ref{eq:proposed_overhead}), are
 $|T|H_k(T)+O(l\sigma b),$
 where $l\le\sigma^k$ is the number of distinct contexts
 $W\!\in\!\Sigma^k$ \!in $T$.
\end{theo}

\begin{IEEEproof}
 See Appendix \ref{sec:CB-proof}.
\end{IEEEproof}
\subsection{Processing time of suffix range queries}
\label{sec:theory:q2}
To evaluate whether Algorithm~\ref{algo:encoded fmsearch} is faster than Algorithm~\ref{algo:fmsearch}, we focus on the time complexity of the rank operation.
As stated in Theorem~\ref{theo:rank-huff-wt}, $rank_w(S,j)$ runs in $O(1+H_0(S))$ time\footnote{\small To be exact, this complexity is proportional to $b$ because practical RRR \cite{PracticalRRR} runs bit-wise rank in $O(b)$ time.}.
Hence, the relationship $H_0(\phi(T_{bwt}))\ll H_0(T_{bwt})$ again explains why \proposedsf is faster than \textsf{ICB-Huff}.
Of course, Algorithm~\ref{algo:encoded fmsearch} incurs an additional cost in calculating $\phi(w|w')$, but this is not serious for a sparse $G_T$.

Moreover, we have the following theorem implying that the search time does not depend on the road network size $\sigma$ but depends only on the maximum out-degree $\delta$ of the road network (which is usually less than four).
\begin{theo}[$\sigma$-independence]
 \label{theo:sigma-scalable}
 Let $P\in E^*$ be any query path (\texttt{\$} is not included).
 Algorithm~\ref{algo:encoded fmsearch} runs in
 $O(|P|\cdot\delta b)$ time.
\end{theo}
\begin{IEEEproof}
 For any $w, w'\in E$,  we have $\eta:=\phi(w|w')\le\delta+2$.
 By the construction of RML, $\eta$ is at least the $\delta+2$-th most frequent symbol in $\phi(T_{bwt})$.
 Thus $\eta$ is at most located at the $\delta+2$ level of the Huffman tree.
 Hence, 
 $rank_{\eta}(\phi(T_{bwt}),j)$ in Eq.~(\ref{eq:pseudo rank}) runs
 in $O(\delta b)$ time (remember the bit-wise rank operation in practical RRR
 \cite{PracticalRRR} requires $O(b)$ time).
 Since \textit{PseudoRank} is calculated at most $2|P|-2$ times in
 Algorithm \ref{algo:encoded fmsearch}, this leads to the conclusion.
\end{IEEEproof}

Other FM-indexes do not satisfy this property.
Note that this time complexity also does not depend on the data size $|T|$.

\subsection{Comparison of RML with \LandC}
\label{sec:theory:landc}
\begin{figure}[b]
 \centering
 \includegraphics[scale=0.7]{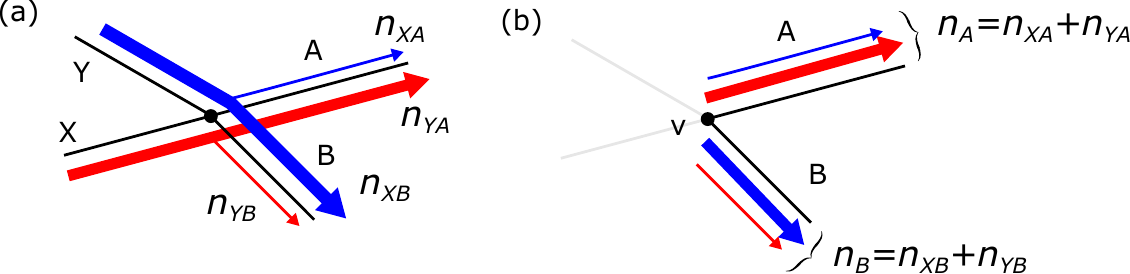}
 \caption{Comparing two NCT labeling methods: (a) RML; and (b) \LandC
 \cite{COMPRESS}}
 \label{fig:LandC}
\end{figure}

Minimum entropy labeling (\LandC) is a labeling scheme for NCTs that was recently proposed in \cite{COMPRESS}, which works as a preprocessor for general compressors, such as Huffman coding or LZ coding (i.e., pattern matching was not considered).
Similar to RML, \LandC converts a sequence of road edges to a low entropy sequence of small integers as follows:
\begin{align}
 w_1w_2\cdots w_n\to \psi(w_1)\psi(w_2)\cdots\psi(w_n)
\end{align}
where $\psi: E\to \mathbb{N}$ is the \LandC function.
Different labels are assigned to road segments that shares head node $v$ (Fig.~\ref{fig:LandC}(b)).
By contrast, our RML conversion is as follows:
\begin{align}
 w_1w_2\cdots w_n\to \phi(w_1|\$)\phi(w_2|w_1)\cdots\phi(w_n|w_{n-1}).
\end{align}
Unlike RML, the \LandC function $\psi$ does not consider the previous symbol.
Specifically, 
\LandC labels based on the \emph{unigram} frequencies, which are shown as $n_\textsf{A}$ and $n_\textsf{B}$ in Fig.~\ref{fig:LandC}(b).
Conversely, our RML, shown in Fig.~\ref{fig:LandC}(a), is based on bigram frequencies, $n_\textsf{XA}$, $n_\textsf{XB}$, $n_\textsf{YA}$, and $n_\textsf{YB}$.

Given these differences, the advantage of RML can be intuitively explained as follows.
Real trajectories tend to go straight rather than turn left or right, as shown in Fig.~\ref{fig:LandC}.
Because RML considers the previous road segment, RML can take account the direction of the movement, whereas such information is lost in \LandC.
This implies that RML can capture a higher-order correlation compared to \LandC.
Although \LandC also has the optimality of entropy, it cannot be better than RML.
The experimental comparison is shown in Section \ref{sec:exp:labeling}.
Mathematically, we have the following theorem.
\begin{theo}
 For any trajectory string $T$, RML achieves a smaller 0th order empirical entropy than MEL does.
\end{theo}
\begin{IEEEproof}
Considering the size of the feasible labeling space, we find that our labeling space $\{\phi(w|w')\}$ is a superset of that of \LandC, $\{\psi(w)\}$.
In other words, \LandC can be emulated by an RML $\bar\phi$ that is not necessarily optimal.
Therefore, the optimality of RML (Theorem \ref{theo:optimality}) leads to the conclusion.
\end{IEEEproof}

\section{Experiments}
\label{sec:experiment}
\subsection{Experimental setup}
\label{sec:exp-setup}
\mypara{Implementation}
All methods were implemented in C{\small++}
and compiled with g{\scriptsize++} (version 4.8.4)
with the \texttt{-O3} option.
We used the
\texttt{sdsl-lite} 
library (version 2.0.1) for (in-memory) wavelet trees ({\small\url{http://github.com/simongog/sdsl-lite/}}).
The BWT was calculated using
\texttt{sais.hxx} ({\small\url{http://sites.google.com/site/yuta256/sais/}}). 
Experiments were conducted 
on a workstation with the following specifications:
Intel Core i7-K5930 3.5GHz CPU (64-bit, 12 cores,
L1 64kB$\times$12, L2 256kB$\times$12, L3 15MB),
DDR4 32GB RAM,
Ubuntu Linux 14.04.
\mypara{Competitors}
Table~\ref{table:methods} lists the competitors used in this paper.
We used five FM-index variants:
uncompressed (\textsf{UFMI}, \textsf{FM-GMR})
and compressed (\textsf{ICB-WM}, \textsf{ICB-Huff}, \textsf{FM-AP-HYB}).
The block-size parameter $b$ had to be specified for \proposedsf, \textsf{ICB-Huff},
and \textsf{ICB-WM}.
Unless otherwise noted, we use $b=63$.
\textsf{FM-GMR} \cite{FM-GMR} and \textsf{FM-AP-HYB} \cite{FM-AP}
are FM-index variants that are tailored for huge $\sigma$
and that support $O(\log\log\sigma)$ rank operation
(faster than the $O(\log\sigma)$ of \textsf{UFMI});
they are available in \texttt{sdsl-lite} library.
These were the fastest (\textsf{FM-GMR})
and the smallest (\textsf{FM-AP-HYB})
methods for huge $\sigma$ in a recent benchmark \cite{CSA++}.

There are many possibilities for compressing NCTs by combining simple techniques such as run-length encoding.
However, we do not consider such techniques in this study because
pattern matching is not supported in sublinear time.
In our prior evaluation, the Boyer-Moore method (linear time search) was at least four orders of magnitude slower than \proposedsf even if $T$ was stored in an in-memory uncompressed array.
In this study, we thus only consider \textsf{RePair} \cite{repair}, a standard benchmark in stringology which showed the best compression ratio in the initial evaluation, and \textsf{PRESS} \cite{PRESS}, which is the shortest-path-based NCT compressor (Note that `Q?' is unchecked in Table~\ref{table:methods} for these methods).

%
%
%
%
\mypara{Measurement}
The search time was averaged over 500 suffix range queries of length 20 randomly sampled from the data.
For evaluation of the data size of \proposedsf, the size of the ET-graph is included.
\begin{figure*}[tb]
 \vspace{-1.1em}
 \begin{center}
  \includegraphics[scale=0.59,trim=0 0 0pt 0]{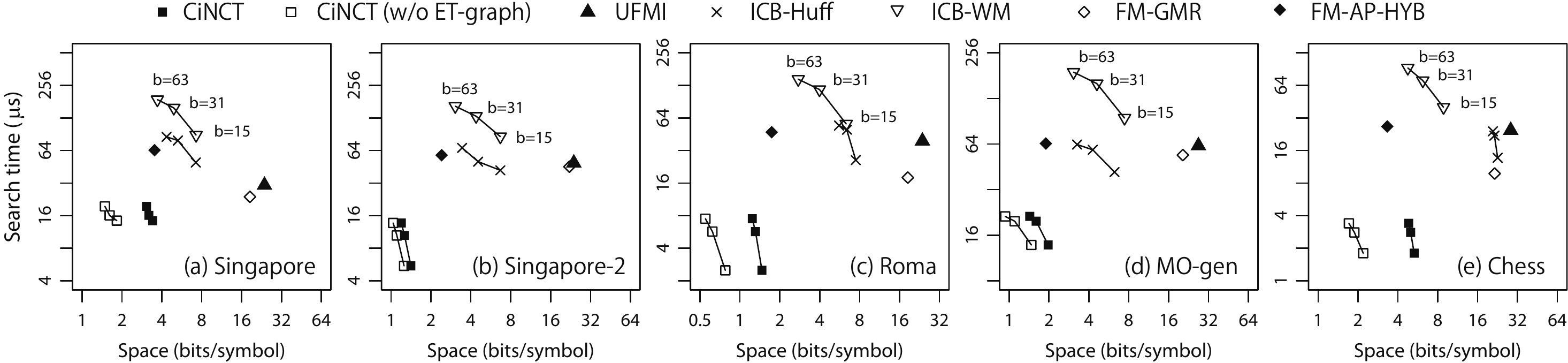}
  \vspace{-0.5em}
  \caption{Data size/search time (suffix range query):
  \textnormal{the proposed method shows the best performance.
  \textsf{\proposed (w/o ET-graph)} is used to show the data size without ET-graph.
  The block size used in the RRR bit vectors is parameterized as $b\in\{15, 31, 63\}$.
  Results for the other methods in Table \ref{table:methods} were omitted
  because their linear-time search was
  too slow.
  }}\label{fig:q1}
  \vspace{-1.5em}
 \end{center}
\end{figure*}
\mypara{Datasets}
The datasets used in this study are as follows:

\begin{itemize}
 \item \textsf{Singapore}: NCTs of taxi cabs used in \cite{PRESS}. 
       This dataset contains many transitions without physical connection.
 \item \textsf{Singapore-2}:
       Preprocessed \textsf{Singapore} dataset such that transitions between two road segments without a physical connection are interpolated with the shortest path.
       This is used to evaluate the gain against the noisy dataset (\textsf{Singapore}).
 \item \textsf{Roma}: GPS trajectories of taxi cabs in Rome. 
       NCT representations were obtained by HMM map-matching \cite{HM4} (\url{http://crawdad.org/roma/taxi/}).
 \item \textsf{MO-gen}:
       NCTs generated by the moving object generator (\url{http://iapg.jade-hs.de/personen/brinkhoff/generator/}). 
 \item \textsf{Chess}:
       All chess game records (Blitz, 2006--2015, 1.87 million games, \url{http://www.ficsgames.org}). Each openings (10 moves) is converted into a string of hash values of Forsyth-Edwards notation.
\end{itemize}
%
%

Although \textsf{Chess} is not a vehicular dataset,
it also has a sparse ET-graph $G_T$ because of the characteristics
of chess games.
This is included to show the possibility that \proposedsf is applicable to
targets other than NCTs.
Table~\ref{table:dataset-stat} lists the statistics of
the datasets, which are used to explain the results.
\begin{table}[tb]
 \begin{center}
  \caption{Our proposed method and its competitors$^\ast$}\label{table:methods}
  \footnotesize
  \begin{threeparttable}
  \begin{tabular}{|p{1.2cm}|l|p{3.28cm}|c|c|} \hline
   \textbf{Method}   & \textbf{Data} & \textbf{Description}
   & \textbf{C?}$^\dagger$ & \textbf{Q?}$^\ddagger$\\ \hline
   \proposedsf
   & $\phi(T_{bwt})$
   & HWT with RRR
   & \checkmark
   & \checkmark
   \\ \hline
   \textsf{UFMI}
   & $T_{bwt}$
   & WM$^\diamond$ \cite{WM} with uncompressed bitmap
	   \cite{UncompressedBitVector}
   &
   & \checkmark  \\ \hline
   \textsf{ICB-WM}
   & $T_{bwt}$
   & WM with RRR \cite{WM}
   & \checkmark
   & \checkmark
       \\ \hline
   \textsf{ICB-Huff}
   & $T_{bwt}$
   & HWT with RRR \cite{ImplicitCompBoost}
   & \checkmark
   & \checkmark
       \\ \hline
   \textsf{FM-GMR}
   & $T_{bwt}$
   & FM-index for huge $\sigma$
     with $O(\log\log\sigma)$ rank \cite{FM-GMR}
   & 
   & \checkmark
       \\ \hline
   \textsf{FM-AP-HYB}
   & $T_{bwt}$
   & FM-index for huge $\sigma$
     with $O(\log\log\sigma)$ rank \cite{FM-AP}
   & \checkmark
   & \checkmark
       \\ \hline
   \textsf{PRESS} \cite{PRESS}
   & $T$
   & The state-of-the-art trajectory compressor
   & \checkmark
   &
   \\ \hline
   \textsf{Re-Pair} \cite{repair}
   & $T$
   & A standard benchmark compressor in stringology
   & \checkmark
   &
   \\ \hline
  \end{tabular}
 \begin{tablenotes}
  \scriptsize
  \item[$^\ast$] For the first four method, the type of WT used is in
  description
  \item[$\dagger$] Uncompressed or compressed / $^\ddagger$Supports suffix range query or not
  \item[$^\diamond$] WM: wavelet matrix
 \end{tablenotes}
 \end{threeparttable}
 \end{center}
\end{table}
\begin{table}[tb]
\begin{center}
 \vspace{-5pt}
 \caption{Statistics of each dataset}
 \label{table:dataset-stat}
 \begin{threeparttable}
 \footnotesize
 \begin{tabular}{|l|r|r|r|r|r|r|} \hline
  \textbf{Dataset}   & $|T|$ & $\lg\sigma$ & $H_0(T)$
  & $H_0(\phi)$\tnote{$\dagger$} & $H_1(T)$ & $\bar d$\;\tnote{$\ddagger$} \\ \hline
  \textsf{Singapore}   & 53M   & 15.5 & 13.8 & 1.8 & 1.5 & 26.8\\
  \textsf{Singapore-2} & 75M   & 15.5 & 14.0 & 1.3 & 1.1 & 4.0 \\
  \textsf{Roma}        & 12M   & 15.5 & 13.0 & 0.9 & 0.7 & 2.4 \\
  \textsf{MO-Gen}      & 193M  & 17.4 & 13.0 & 2.8 & 2.5 & 8.8 \\
  \textsf{Chess}       & 20M   & 18.8 & 10.3 & 2.0 & 1.4 & 1.6 \\ \hline
 \end{tabular}
 \begin{tablenotes}
  \scriptsize
  \item[$\dagger$] $H_0(\phi)$ means $H_0(\phi(T_{bwt}))$
  \item[$\ddagger$] $\bar d$ is the average out-degree of the ET-graph $G_T$. \end{tablenotes}
 \end{threeparttable}
 \vspace{-1.5em}
\end{center}
\end{table}
%
%
\subsection{Comparison with various FM-indexes}\label{sec:q1}
Evaluation results for data size and processing time of suffix range queries are shown in Fig.~\ref{fig:q1}.
%
We observe that \proposedsf requires \emph{less than 2 bits per symbol} to store NCTs, and pattern matching of length 20 is processed in \emph{a few tens of microseconds}.
We also observe that \emph{\proposedsf outperforms the competitors in terms of both data size and query processing time.}
We explain these results in detail below.
\mypara{Data size}
Compared with \textsf{ICB-Huff} and \textsf{ICB-WM}, \proposedsf reduces the data size by up to 78\% and 57\%, respectively.
%
As explained in Section~\ref{sec:theory}, the space overhead decreases if $H_0(S)$ decreases.
From Table~\ref{table:dataset-stat} we can confirm that $H_0(\phi(T_{bwt}))\ll H_0(T_{bwt})$
holds for all datasets (note that $H_0(T)=H_0(T_{bwt})$).
This explains why \proposedsf shows this significant improvement.
\begin{figure*}[tb]
 \centering
  \hspace{-1em}
 \begin{minipage}[b]{.2\textwidth}
  \includegraphics[scale=.56]{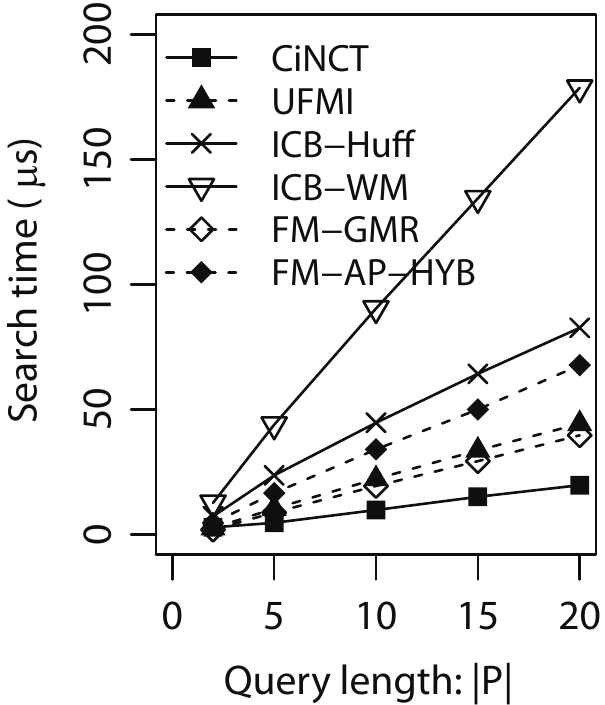}
  \caption{$|P|$ vs. search time: \textnormal{(\textsf{Singapore}
  dataset)}}
  \label{fig:querylen}
 \end{minipage}
 \hspace{0.8em}
 \begin{minipage}[b]{.36\textwidth}
  \includegraphics[scale=.6]{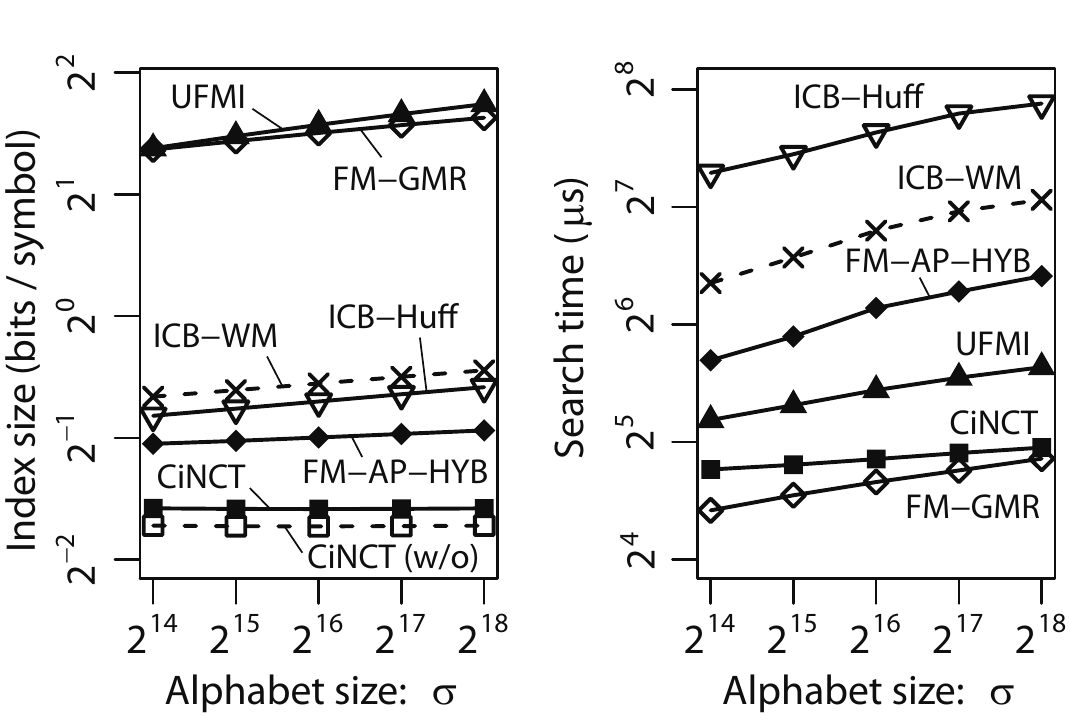}
  \caption{\proposedsf shows the best $\sigma$-dependence
 \textnormal{(Left: index size, right: search time
 / \textsf{RandWalk} dataset)}}
  \label{fig:q6}
 \end{minipage}
 \hspace{1em}
 \begin{minipage}[b]{.36\textwidth}
  \includegraphics[scale=.6]{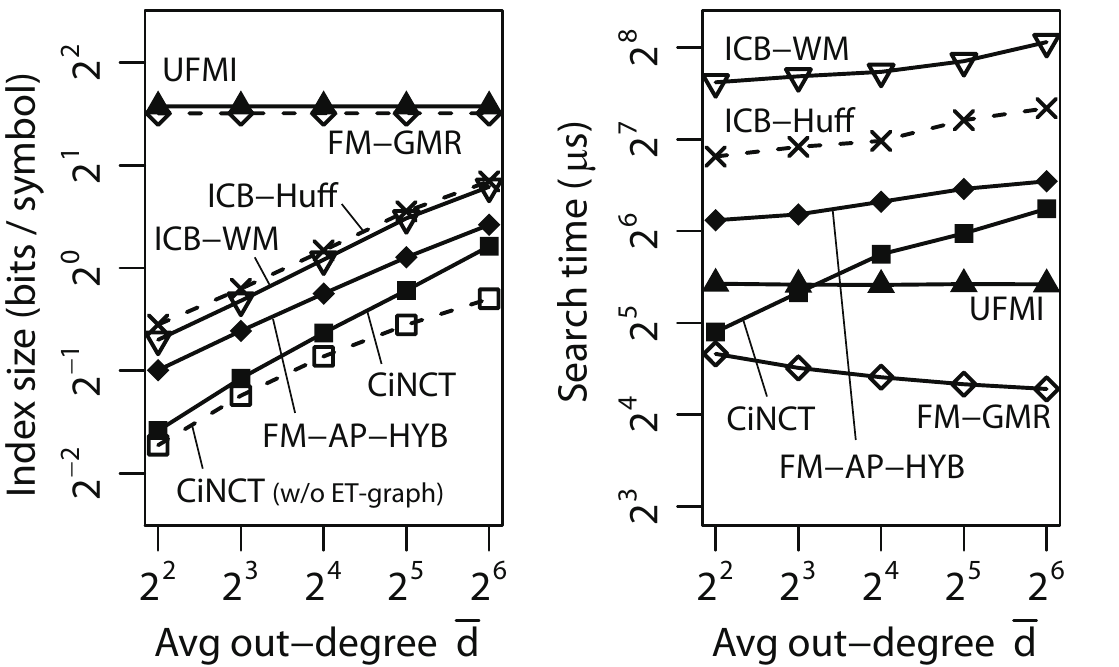}
  \caption{Dependence on out-degree \textnormal{(Left: index size, right:
  search time / \textsf{RandWalk} dataset)}}
  \label{fig:out-degree-dep}
 \end{minipage}
\end{figure*}
\hspace{-7pt}
%
\proposedsf even shows better compression than
the smallest variant \textsf{FM-AP-HYB}, which was designed for huge $\sigma$.
The improvement in \textsf{Singapore-2} is larger than that of \textsf{Singapore}.
Because ``gapped'' transitions are interpolated in \textsf{Singapore-2},
the ET-graph gets sparser ($\bar{d}=\!26.8\!\to\!4$ in Table~\ref{table:dataset-stat}). This reduces the overhead regarding
the ET-graph (this is confirmed through the difference of \proposedsf and \proposedsf \textsf{(w/o ET-graph)}).
%
\mypara{Processing time of suffix range queries}
\proposedsf is always much faster than \textsf{ICB-Huff} and \textsf{ICB-WM};
the speedups are up to 7 and 25 times, respectively.
Surprisingly, \proposedsf is even faster than those of the uncompressed indexes (\textsf{UFMI} and \textsf{FM-GMR}).
Again, this speedup can be explained by the shallowness of the HWT of \proposedsf.
Because we have $H_0(\phi(T_{bwt}))\ll H_0(T_{bwt})$, the HWT becomes shallower. 
This decreases the number of bit-wise \textit{rank} operations in the HWT, as pointed out in Section~\ref{sec:theory:q2}.

%
\mypara{Effect of block size $b$}
As mentioned in Section~\ref{sec:trajwt}, as $b$ becomes larger, the results show better compression but slower search.
However, as shown in Fig.~\ref{fig:q1}, \emph{the sensitivity to the block size parameter $b$ is very small for \proposedsf.}
This indicates that the proposed method is nearly \emph{parameter-free}.

\mypara{Effect of $|P|$}
Figure~\ref{fig:querylen} shows how the processing time of suffix range queries increases as the query length $|P|$ increases. 
%
For all methods, the processing time grows linearly as the query length increases, because the numbers of iterations in Algorithm~\ref{algo:fmsearch} and Algorithm~\ref{algo:encoded fmsearch} is proportional to $|P|$.
We observe that \emph{\proposedsf shows the slowest growth among all methods.}

\subsection{Comparison with several compression methods}\label{sec:q2}
Table~\ref{table:q2} compares the compression ratio, which is defined as the uncompressed size (binary file of 32-bit integers) divided by the compressed size.
%
%
%
%
%
We observe that \proposedsf shows better compression than the existing methods.
In particular, \emph{our method is better than MEL, which also showed the best compressibility in recent evaluation of NCT compression} \cite{COMPRESS}.
Note that the road network storage is not included in MEL evaluations whereas it is considered for \proposedsf (as ET-graph).
%

\begin{table}
 \begin{center}
  \caption{Compression ratio (larger is better)}
  \label{table:q2}
  \vspace{-0.2em}
 \begin{threeparttable}
  \footnotesize
  \begin{tabular}{|l|r|r|r|r|r|}\hline
                   & \multicolumn{1}{p{1.2cm}|}{\textsf{Singapore}}
                   & \multicolumn{1}{p{1.55cm}|}{\textsf{Singapore-2}}
                   & \multicolumn{1}{p{.7cm}|}{\textsf{Roma}}
                   & \multicolumn{1}{p{1.1cm}|}{\textsf{Mo-Gen}}
                   & \multicolumn{1}{p{.7cm}|}{\textsf{Chess}} \\\hline
   \proposedsf     & \textbf{10.5}& \textbf{27.0}&\textbf{25.2}& \textbf{25.6}& 10.3\\
   \textsf{MEL}$^\dagger$    & \textsc{n/a}& 15.8& 21.2& \textsc{n/a}& \textsc{n/a}\\
   \textsf{Re-Pair}& 8.4& 11.4& 20.6& 20.6& \textbf{11.0}\\
   \textsf{bzip2}  & 5.3&  5.6& 13.6& 5.3& 7.1\\
   \textsf{PRESS}$^\ddagger$  & 4.6& \textsc{n/a}&\textsc{n/a}& \textsc{n/a}& \textsc{n/a}\\
   \textsf{zip}    & 2.5& 2.5& 5.0& 2.6& 3.9\\ \hline
  \end{tabular}
 \begin{tablenotes}
  \scriptsize
  \item[$\dagger$] Huffman coding was used after labeling, as in \cite{COMPRESS}. We evaluated only for ungapped datasets because MEL assumes no gap (see \textsf{Singapore-2} explanation in Sec.\ref{sec:exp:dataset}).
  \item[$\ddagger$] Only the result for the \textsf{Singapore} dataset
  \cite{PRESS} is shown because no available implementation was found.
 \end{tablenotes}
 \end{threeparttable}
 \end{center}
\end{table}
\subsection{Effect of labeling strategy}\label{sec:exp:labeling}
\mypara{Comparison with \LandC}
According to our analysis in Section \ref{sec:theory:landc}, RML achieves lower entropy than \LandC does.
We show our experimental results from two real NCT datasets, i.e., \textsf{Singapore2} and \textsf{Roma}.
Table \ref{table:LandC} provides a comparison of the entropy achieved by RML and \LandC.
These results show that our RML obtained approximately a 30\% smaller entropy than that of \LandC.
\begin{table}[bt]
 \centering
 \caption{Comparison of entropy (RML and \LandC)}
 \vspace{-.5em}
 \label{table:LandC}
 \begin{tabular}{|l|r|r|}\hline
  Dataset             & RML (Proposed)& \LandC \cite{COMPRESS}\\\hline
  \textsf{Singapore2} & \textbf{1.26}& 1.93\\
  \textsf{Roma}       & \textbf{0.76}& 0.99\\\hline
 \end{tabular}
\end{table}
\subsubsection{Optimality}
\begin{figure}[tb]
 \centering
  \includegraphics[scale=.55]{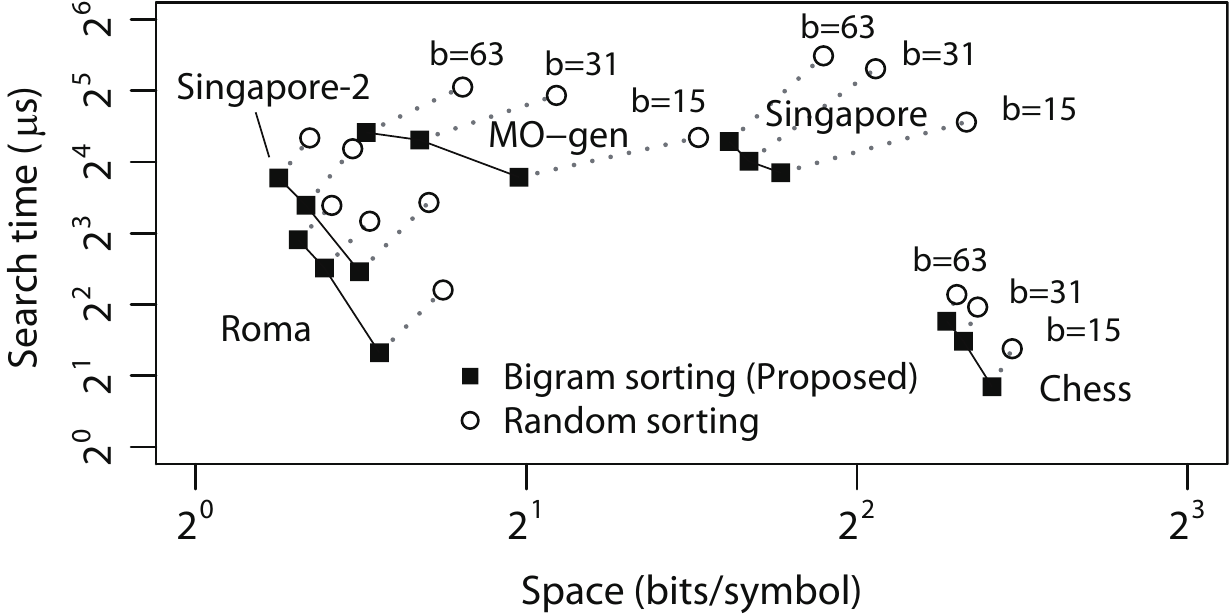}
  \caption{Comparison of labeling strategies}
  \label{fig:rand-sort}
\end{figure}

In Section~\ref{sec:traj encoding}, we proposed a labeling strategy that assigns small integers $c_{ww'}$ sorted by the bigram counts $n_{ww'}$.
The data size and search time under this strategy are expected to be better than those of any other possible labeling strategy, because we showed the optimality of our strategy (Theorem~\ref{theo:optimality}).
Here, we compare our strategy with the \textit{random sorting strategy}, which assigns randomly shuffled small integers
$c_{ww'}\in \{1,\cdots, |N_{out}(w')|\}$.
Figure~\ref{fig:rand-sort} shows the comparison for the five datasets ($b\in\{15,31,63\}$).
We observe that \emph{the index size and the search time of the bigram sorting strategy are always better than those of random sorting strategy.}
Compared to the random strategy, it reduces the data size by up to 32\%, and the search time by up to 57\%.
These results indicate the importance of the bigram sorting strategy.
\subsection{Effect of ET-graph size/shape}
\label{sec:sigma-scalable}
\mypara{Effect of $\sigma$}
In Theorem~\ref{theo:sigma-scalable}, we showed that the search time of \proposedsf does not depend on the size $\sigma$ of the road map.
Here, we investigate what happens when $\sigma$ grows.
For the experiment, we use synthetic data \textsf{RandWalk}: random walks on a directed random Poisson graph.
The average out-degree $\bar d$ of the graphs is fixed at four, and $|T|$ is set to $800\sigma$.

In Fig.~\ref{fig:q6}, \emph{\proposedsf shows good scalability against $\sigma$, whereas the index sizes and the search times of the existing methods both increase.}
The search time of \proposedsf is almost constant,
as predicted by Theorem~\ref{theo:sigma-scalable}.\footnote{\small In fact,
the search time of \proposedsf increases slightly
because $|T|$ increases with $\sigma$ in our setting, leading to a lower
cache hit ratio. We confirmed the exact
constant search time when $|T|$ is constant.}
The other methods do not show this property.
For example, both the index size and the search time of \textsf{UFMI} at
$\sigma=2^{18}$ are 30\% larger compared to the $\sigma=2^{14}$ case.
%
\mypara{Effect of sparsity}
Next, we investigate how the data size and search time behave against the average out-degree $\bar d$.
Figure~\ref{fig:out-degree-dep} shows the results for the \textsf{RandWalk} dataset used in Section~\ref{sec:sigma-scalable}.
For comparison, we fixed $\sigma=2^{16}$ and $|T|=100$M, and
changed $\bar d$ between $2^2$ and $2^7$.
We observe that \emph{the sparsity of the ET-graph is the key factor for \proposedsf.}
Although the compression performance of \proposedsf is the best, the data size grows quickly.
This is due to two factors: the increase of ET-graph size and the increase of the depth of HWT.
This result is a natural consequence of our assumption that the road network is highly sparse.
However, this result shows that our method works for larger $\bar d$ than in the road network case, $\bar d\simeq 2^2$.
This result opens the door to applications to datasets not mentioned in this paper (e.g., symbol-valued time series).
\subsection{Sub-path extraction time}
\begin{figure}[tb]
 \begin{center}
  \hspace{-1em}
 \begin{minipage}[b]{.22\textwidth}
  \includegraphics[scale=.55,trim=3pt 0pt 0 0pt]{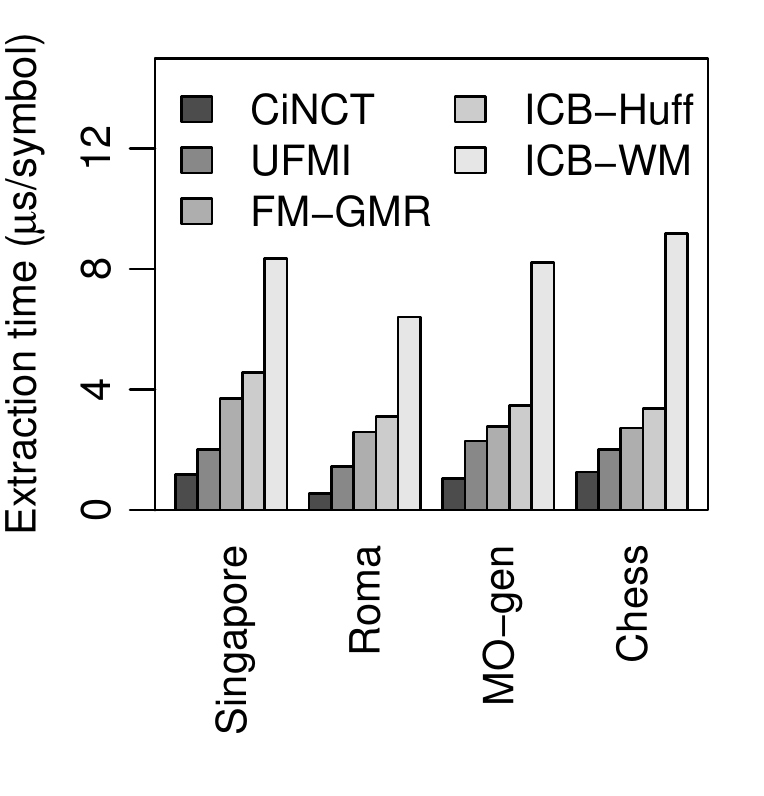}
  \vspace{-2em}
  \caption{Extraction time\vspace{1em}}\label{fig:extraction}
 \end{minipage}
 \hspace{0.5em}
 \begin{minipage}[b]{.22\textwidth}
  \includegraphics[scale=.55,trim=3pt 0pt 0pt 0pt]{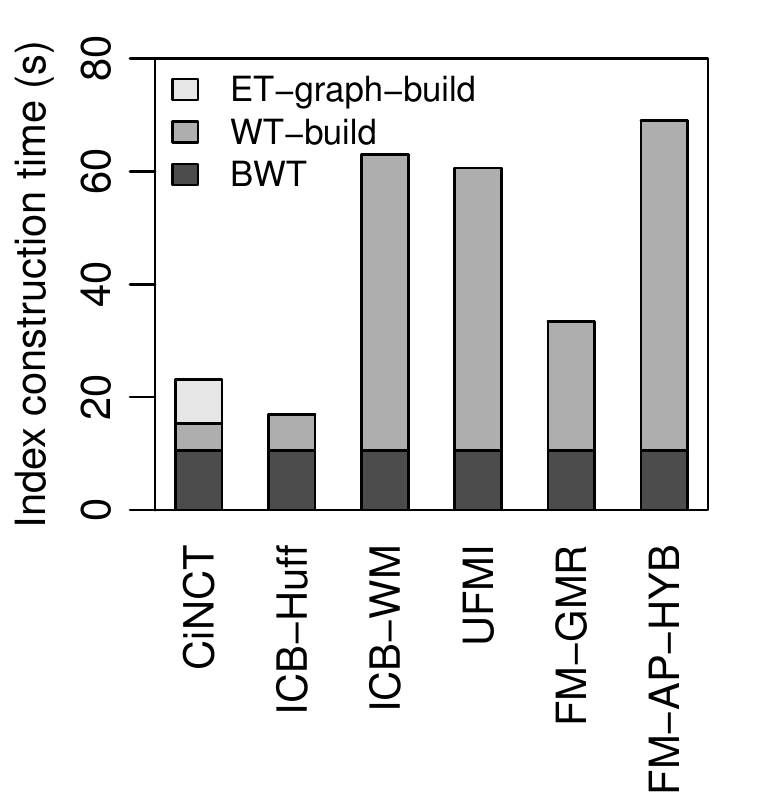}
  \vspace{-2em}
  \caption{Index construction time \textnormal{(\textsf{Singapore} dataset)}}\label{fig:q5}
 \end{minipage}
 \end{center}
 \vspace{-2em}
\end{figure}

Here, we evaluate \emph{extract} queries described in Section~\ref{sec:extract}.
We evaluated the extraction time for
obtaining the entire $T$, that is, $l=|T|$ and $j=0$.
Figure~\ref{fig:extraction} compares the extraction times (per symbol) for the four datasets.
We observe that \emph{\proposedsf shows the fastest extraction among the competitors
} (twice as fast as \textsf{UFMI}).
Again, this can be explained by the fast \emph{rank} calculation in \proposedsf
(\textit{PseudoRank}), as discussed above.
Note that we omitted the results for \textsf{FM-AP-HYB} because
random access to $T_{bwt}$ was not supported in the \texttt{sdsl-lite} library.

\subsection{Index construction time}
Figure~\ref{fig:q5} compares the index construction times of FM-index variants.
\emph{The construction time of \proposedsf is comparable to
that of \textsf{ICB-Huff}, and shorter than those of the other methods.}
In fact, \proposedsf is the second fastest among the considered competitors.
\textsf{ET-graph-build} in Fig.~\ref{fig:q5}
includes all operations that are not needed for the other methods.
Here, we can see the overhead for the construction of the ET-graph
is not a serious problem.
Note that all additional operations, including the construction of $G_T$ from $T$, obtaining RML function $\phi$,
labeling $T_{bwt}$, and calculation of $Z_{w'w}$, can be executed in linear time $O(|T|)$, which implies the scalability
of construction.

\section{Related work}
\label{sec:related-work}
\mypara{Trajectory indexing}
One important application of our method is trajectory indexing.
Although there are numerous studies on this topic as shown in a survey \cite{STSurvay}, we present only the most relevant ones here.
MON-tree \cite{MON-tree} is one of the most famous methods to index NCTs.
This method, as well as many of other NCT indexing methods, mainly focuses on spatio-temporal range queries.
Krogh et al.~\cite{Krogh2014} proposed a data structure similar to MON-tree designed for a different type of query called \emph{strict path query}, which aims to find trajectories in a database that traveled along a given sub-path $P$ during a given time interval $I$.
This method was implemented based on B$^+$-trees.
%
Koide et al.~\cite{SNT-index} showed that strict path queries can be efficiently processed using suffix range queries.
This method is a hybrid data structure that indexes timestamps and spatial paths using B$^+$-trees and FM-index, respectively.
Brisaboa et al.~\cite{CTR} also employed a compressed suffix array to store spatial paths.
As for timestamps, an in-memory structure was also used.
These methods (\cite{CTR} and \cite{SNT-index}) used existing methods to compress the suffix arrays.
Our method can be regarded as one that boosts such methods in terms of memory storage and query processing time.
%
%
\mypara{Trajectory compression}
As noted in Section \ref{sec:intro}, shortest-path encodings have been used to compress spatial paths in several papers \cite{COMPRESS, MMTC, SPNET} and \cite{PRESS}.
As implied by Krumm \cite{PPP}, an NCT dataset is expected to have a small $k$-th order empirical entropy (Eq.~(\ref{eq:Hk})), but none of such shortest-path-based compressors have provided an information-theoretic evaluation of the compressed size.
As an NCT compressor, our method first focuses on high-order entropy and gives an information-theoretic bound (Theorem~\ref{theo:proposedCB}).
One of the methods proposed in \cite{COMPRESS}, \LandC, is a different type of spatial path compressor that achieves higher compressibility than shortest-path-based methods.
As shown in Section \ref{sec:theory:landc}, RML achieves a smaller entropy than \LandC does.
In \cite{PopaSTC}, graph partitioning was used to reduce the size of spatial paths.

To compress timestamps in NCTs, lossy compression methods are used in \cite{COMPRESS, SPNET, PopaSTC} and \cite{PRESS}, whereas lossless compression was used in \cite{CTR}.
These NCT compressors support some useful queries.
In any case, timestamps should be stored and indexed to fit the purpose of use.
In order to realize such queries in a smaller storage, it is an interesting research direction to combine these timestamp compressors with \proposedsf.
\mypara{FM-index}
FM-index, a compressed representation of suffix arrays \cite{SA}, was proposed by Ferragina and Manzini \cite{FM-index}.
We have already described FM-index and the related topics in Section \ref{sec:preliminaries}.
Although there are a number of FM-index variants (e.g., \cite{FCBC, ImplicitCompBoost, FM-GMR, FM-AP}), these are essentially designed for general strings.
In Section \ref{sec:experiment}, we compared our method also with FM-indexes designed for large alphabet \cite{FM-GMR, FM-AP}.
For large $\sigma$, these methods can process suffix range queries in $O(|P|\log\log\sigma)$ time, which is much faster than typical $O(|P|\log\sigma)$ time.
Importantly, we employed the domain-specific knowledge of the target data (i.e., sparse transition in road networks) to enhance the compression and query processing.
This point is the largest difference from the FM-index family designed for general strings.

\section{Conclusion}
\label{sec:conclusion}

In this paper, we proposed \proposedsf, a novel compressed data structure capable of handling a very large number of NCTs.
We incorporated the sparsity of road networks into the FM-index by using our proposed RML and \emph{PseudoRank} techniques.
The resulting data structure supports pattern matching (i.e., via suffix range queries) and sub-path extraction from an arbitrary position while still achieving high compressibility.
Our experiments showed that \proposedsf outperformed existing methods in terms of index size and search time, as shown above Fig.~\!\ref{fig:q1}, Table~\ref{table:q2}, and Fig.~\!\ref{fig:q6}.
Our method was even faster than an uncompressed index.
We also discussed theoretically why \proposedsf is compact and fast.
Further, we proved the optimality of RML, i.e., the smallest size and the fastest search are achieved.
We also showed that RML performed even better than the state-of-the-art NCT labeling method (\LandC).

Our data structure has a wide range of applications in which pattern matching based on spatial paths is a key component.
In fact, our method can be directly applied to some pioneering methods for spatio-temporal NCT processing \cite{CTR, SNT-index}.
Given our method, we also expect that a practical spatio-temporal database system will become possible in the future.

\newpage
\bibliographystyle{IEEEtran}
\bibliography{fmindex}

\begin{thebibliography}{10}
\providecommand{\url}[1]{#1}
\csname url@samestyle\endcsname
\providecommand{\newblock}{\relax}
\providecommand{\bibinfo}[2]{#2}
\providecommand{\BIBentrySTDinterwordspacing}{\spaceskip=0pt\relax}
\providecommand{\BIBentryALTinterwordstretchfactor}{4}
\providecommand{\BIBentryALTinterwordspacing}{\spaceskip=\fontdimen2\font plus
\BIBentryALTinterwordstretchfactor\fontdimen3\font minus
  \fontdimen4\font\relax}
\providecommand{\BIBforeignlanguage}[2]{{%
\expandafter\ifx\csname l@#1\endcsname\relax
\typeout{** WARNING: IEEEtran.bst: No hyphenation pattern has been}%
\typeout{** loaded for the language `#1'. Using the pattern for}%
\typeout{** the default language instead.}%
\else
\language=\csname l@#1\endcsname
\fi
#2}}
\providecommand{\BIBdecl}{\relax}
\BIBdecl

\bibitem{COMPRESS}
Y.~Han, W.~Sun, and B.~Zheng, ``{COMPRESS}: A comprehensive framework of
  trajectory compression in road networks,'' \emph{ACM Trans. Database Syst.},
  vol.~42, no.~2, pp. 11:1--11:49, May 2017.

\bibitem{MMTC}
G.~Kellaris, N.~Pelekis, and Y.~Theodoridis, ``Map-matched trajectory
  compression,'' \emph{J. Syst. Softw.}, vol.~86, no.~6, pp. 1566--1579, 2013.

\bibitem{CTR}
N.~R. Brisaboa, A.~Fari{\~{n}}a, D.~Galaktionov, and M.~A. Rodr{\'i}guez,
  ``Compact trip representation over networks,'' in \emph{Proc. SPIRE'16},
  2016, pp. 240--253.

\bibitem{SPNET}
B.~Krogh, C.~S. Jensen, and K.~Torp, ``Efficient in-memory indexing of
  network-constrained trajectories,'' in \emph{Proc. GIS '16}, no.~17, 2016.

\bibitem{PopaSTC}
I.~{Sandu Popa}, K.~Zeitouni, V.~Oria, and A.~Kharrat, ``{Spatio-temporal
  compression of trajectories in road networks},'' \emph{GeoInformatica},
  vol.~19, no.~1, pp. 117--145, 2014.

\bibitem{SNT-index}
S.~Koide, Y.~Tadokoro, and T.~Yoshimura, ``{SNT-index: Spatio-temporal index
  for vehicular trajectories on a road network based on substring matching},''
  in \emph{Proc. SIGSPATIAL UrbanGIS Workshop'15}, 2015.

\bibitem{SA}
U.~Manber and G.~Myers, ``Suffix arrays: a new method for on-line string
  searches,'' in \emph{Proc. SODA'90}, 1990.

\bibitem{FM-index}
P.~Ferragina and G.~Manzini, ``Opportunistic data structures with
  applications,'' in \emph{Proc. FOCS'00}, 2000.

\bibitem{BWT}
M.~Burrows and D.~J. Wheeler, ``A block-sorting lossless data compression
  algorithm,'' in \emph{Technical Report 124}.

\bibitem{WT}
R.~Grossi, A.~Gupta, and J.~S. Vitter, ``High-order entropy-compressed text
  indexes,'' in \emph{Proc. SODA'03}, 2003, pp. 841--850.

\bibitem{UncompressedBitVector}
G.~Jacobson, ``Space efficient static trees and graphs,'' in \emph{Proc.
  FOCS'89}, 1989, pp. 549--554.

\bibitem{RRR}
R.~Raman, V.~Raman, and S.~Rao, ``Succinct indexable dictionaries with
  applications to encoding k-ary trees and multisets,'' in \emph{Proc.
  SODA'02}, 2002, pp. 233--242.

\bibitem{WTAll}
G.~Navarro, ``Wavelet trees for all,'' in \emph{Proc. CPM'12}, 2012, pp. 2--26.

\bibitem{HuffmanWT}
V.~M\"{a}kinen and G.~Navarro, ``New search algorithm and time/space tradeoffs
  for succinct suffix array,'' in \emph{Tech.Rep.~\!C-2004-20, Univ.\! of
  Helsinki}, 2004.

\bibitem{EmpiricalEntropy}
G.~Manzini, ``An analysis of the {Burrows-Wheeler} transform,'' \emph{J. ACM},
  vol.~48, no.~3, pp. 407--430, 2001.

\bibitem{FCBC}
J.~K{\"a}rkk{\"a}inen and S.~J. Puglisi, ``{Fixed block compression boosting in
  FM-Indexes},'' in \emph{Proc. SPIRE'12}, 2011, pp. 174--184.

\bibitem{ImplicitCompBoost}
V.~M\"{a}kinen and G.~Navarro, ``Implicit compression boosting with
  applications to self-indexing,'' in \emph{Proc. SPIRE'07}, 2007, pp.
  229--241.

\bibitem{WM}
F.~Claude and G.~Navarro, ``The wavelet matrix,'' in \emph{Proc. SPIRE'12},
  2012, pp. 167--179.

\bibitem{PracticalRRR}
G.~Navarro and E.~Providel, ``{Fast, small, simple rank / select on bitmaps},''
  in \emph{Proc. SEA'12}, 2012, pp. 295--306.

\bibitem{FM-GMR}
A.~Golynski, J.~I. Munro, and S.~S. Rao, ``Rank/select operations on large
  alphabets: A tool for text indexing,'' in \emph{Proc. SODA'06}, 2006, pp.
  368--373.

\bibitem{FM-AP}
J.~Barbay, T.~Gagie, G.~Navarro, and Y.~Nekrich, ``Alphabet partitioning for
  compressed rank/select and applications,'' in \emph{Proc. ISAAC'10}, 2010,
  pp. 315--326.

\bibitem{CSA++}
S.~Gog, A.~Moffat, and M.~Petri, ``{CSA++:} fast pattern search for large
  alphabets,'' in \emph{Proc. {ALENEX}'17}, 2017, pp. 73--82.

\bibitem{repair}
N.~Larsson and A.~Moffat, ``Offline dictionary-based compression,'' in
  \emph{Proc. DCC '99}, 1999, pp. 296--305.

\bibitem{PRESS}
R.~Song, W.~Sun, B.~Zheng, and Y.~Zheng, ``{PRESS:} {A} novel framework of
  trajectory compression in road networks,'' \emph{{PVLDB}}, vol.~7, no.~9, pp.
  661--672, 2014.

\bibitem{HM4}
P.~Newson and J.~Krumm, ``{Hidden Markov map matching through noise and
  sparseness},'' in \emph{Proc GIS'09}, 2009, pp. 336--343.

\bibitem{STSurvay}
L.~Nguyen{-}Dinh, W.~G. Aref, and M.~F. Mokbel, ``Spatio-temporal access
  methods: Part 2 (2003--2010),'' \emph{{IEEE} Data Eng. Bull.}, vol.~33,
  no.~2, pp. 46--55, 2010.

\bibitem{MON-tree}
V.~T. de~Almeida and R.~H. G\"{u}ting, ``Indexing the trajectories of moving
  objects in networks,'' \emph{Geoinformatica}, vol.~9, no.~1, pp. 33--60,
  2005.

\bibitem{Krogh2014}
B.~Krogh, N.~Pelekis, Y.~Theodoridis, and K.~Torp, ``Path-based queries on
  trajectory data,'' in \emph{Proc. GIS'14}, 2014, pp. 341--350.

\bibitem{PPP}
J.~Krumm, ``{A Markov model for driver turn prediction},'' in \emph{Society of
  Automotive Engineers (SAE) 2008 World Congress}, 2008.

\bibitem{HuffWT-proof}
``On compressing permutations and adaptive sorting,'' \emph{Theoretical
  Computer Science}, vol. 513, pp. 109--123, 2013.

\end{thebibliography}

%



\newpage
\appendix
\subsection{Proof of Theorem \ref{theo:optimality}}
\label{sec:optimality-proof}
To begin with, let us introduce some mathematical notations.
Let us denote a set of integers as $[\sigma]:=\{1,\cdots, \sigma\}$.
Consider $\sigma$ discrete probability distributions
$p_1, \cdots, p_\sigma$ on $[\sigma]$ defined by
\begin{align}
 p_{w'}(w)=\frac{n_{ww'}}{n_{\cdot w'}},
\end{align}
where
$n_{ww'}$ is the number of bigrams $ww'$ in $T$ and
$n_{\cdot w'}=\sum_{w}n_{ww'}$.
First, we define a permutation of a distribution.
\begin{defi}
 Let $p$ be a discrete distribution on $[\sigma]$.
 A permutated distribution $p^{\pi}$ is a distribution
 where $p^\pi(k)=p(\pi(k))$. Here $\pi$ is a permutation on $[\sigma]$.
\end{defi}

In addition, we introduce the concept of a \emph{decreasing distribution}:
\begin{defi}
 A discrete distribution $p$ is decreasing iff $p(w)\ge p(w+1)$ for
 $\forall w\in[\sigma]$.
 Let $\mathcal{F}$ be a set of decreasing distributions and
 $\mathcal{F}^c$ be a set of non-decreasing distributions.
\end{defi}
Note that we can always find a permutation $\pi$ that
makes any distribution $p$ decreasing, that is, $p^\pi\in\mathcal{F}$.

Let us relate the above definitions to our problem.
Since any possible RML corresponds to an assignment of
distinct integers $c_{ww'}\in[\sigma]$ as mentioned in Section
\ref{sec:traj encoding}, we can regard it as
an array of permutations $\Pi=[\pi_1, \cdots, \pi_\sigma]$.
We denote such an encoder as $\phi^{\Pi}$.
Our strategy, sorting by bigram $n_{ww'}$, corresponds to an array of permutations
$\Pi$ such that each $\pi_i$ makes the distribution $p_i$ decreasing.
Note that, if $w\notin N_{out}(w')$,
we can treat such cases as $p_{w'}(w)=0$.

Our problem is to find an encoder $\phi^{\Pi}$ that achieves the minimum
$H_0(\phi^{\Pi}(T_{bwt}))$.
Consider a mixture distribution
\begin{align}
 p^{\Pi}=\sum_{i\in[\sigma]}\alpha_i p_i^{\pi_i}
 \label{eq:theo:mixture}
\end{align}
where $\alpha_i=n_{\cdot i}/\sum_{j}n_{\cdot j}$.
Since the entropy of a discrete distribution is defined as
\begin{align}
 H(p)=-\sum_{k\in[\sigma]}p(k)\lg p(k),
\end{align}
the following equality holds:
\begin{align}
 H(p^\Pi)=H_0(\phi^{\Pi}(T_{bwt})).
\end{align}
Therefore, we can reformulate our optimization problem as follows.
\begin{align}
 \Pi^*=\text{argmin}_\Pi \; H(p^\Pi).
\end{align}
Consider an optimal $\Pi^*$ and any permutation $\pi$.
Permutating elements in $\Pi^*$ by $\pi$ also yields
another optimal solution by definition:
$H(p^{\Pi^*})=H(p^{\pi\circ\Pi^*})$ where
$\pi\circ\Pi^*=\{\pi\circ\pi_1,\cdots,\pi\circ\pi_n\}$.
Here $g\circ f$ indicates a composite function.
We can therefore assume $p^{\Pi^*}$ is a decreasing
distribution without loss of generality.

We now prove the following Theorem that directly
leads to Theorem \ref{theo:optimality}.
\begin{theo}
 \label{theo:optimality-2}
The optimal solution $\Pi^*$
consists of permutations such that each
$\pi_i\in\Pi^*$ makes the distribution $p_i$ decreasing:
$p_i^{\pi_i}\in\mathcal{F}$ for $\forall i\in[\sigma]$.
\end{theo}

We first consider the following Lemma which qualitatively implies
that a more concentrated distribution has smaller entropy.
\begin{lemm}
 \label{theo:lemm-convex}
 If $a> b\ge0$ and $\varepsilon>0$, we have
 \begin{align}
  -a\lg a -(b+\varepsilon)\lg(b+\varepsilon)
  +(a+\varepsilon)\lg(a+\varepsilon)+b\lg b
  > 0.
  \label{eq:theo:optimality-lemm}
 \end{align}
\end{lemm}
\begin{IEEEproof}
 Since $g(x)=(x+\varepsilon)\lg(x+\varepsilon)-x\lg x$
 is a strictly increasing function,
 we have $g(a)-g(b)>0$, which is equivalent to
 Eq.\,\ref{eq:theo:optimality-lemm}.
\end{IEEEproof}

\vspace{1em}
Now we are ready to prove Theorem \ref{theo:optimality-2}.
\begin{IEEEproof}[Proof of Theorem \ref{theo:optimality-2}]
 We prove optimality by contradiction.
 Let $\Pi^+$ be a set of permutations that minimizes $H$.
 As discussed above, we can assume $p^{\Pi^+}\in\mathcal{F}$
 without loss of generality.
 Let us assume that there exists at least one $\pi_i\in\Pi^+$
 such that $p_i^{\pi_i}\in\mathcal{F}^c$.
 Let us define $q:=p_i^{\pi_i}$.

 Since $q\in\mathcal{F}^c$, there exists $k\in[\sigma]$ such that
 $q(k)<q(k+1)$.
 Based on Eq.\,\ref{eq:theo:mixture}, $p^{\Pi^+}$ can be
 decomposed as
 \begin{align}
  p^{\Pi^+}=(1-\alpha_i)\hat p+\alpha_iq
 \end{align}
 where $\hat p:=\frac{1}{1-\alpha_i}\sum_{j\ne i} \alpha_{j}p_{j}^{\pi_{j}}$.
 If $\hat p(k)\le\hat p(k+1)$, we have
 $p^{\Pi^+}(k)<p^{\Pi^+}(k+1)$, which contradicts the assumption
 $p^{\Pi^+}\in\mathcal{F}$.
 Therefore,
 we have $\hat p(k)>\hat p(k+1)$.

 Consider a permutation $s_k$ which only swaps $k$ and $k+1$ and
 a permutated distribution $q^{s_k}$.
 Let us define $\alpha:=\alpha_i$ and $\beta:=1-\alpha_i$.
 We can calculate the difference of entropy functions between
 the optimal solution $p^{\Pi^+}=\beta\hat p+\alpha q$
 and the swapped distribution $\beta\hat p+\alpha q^{s_k}$:
 \begin{align}
  &H(p^{\Pi^+})-H(\beta\hat p+\alpha q^{s_k})=\nonumber\\
  &
  -\{\beta\hat p(k)+\alpha q(k)\}\lg\{\beta\hat
  p(k)+\alpha q(k)\}
  \nonumber\\
  &-\{\beta\hat p(k+1)+\alpha q(k+1)\}\lg\{\beta\hat
  p(k+1)+\alpha q(k+1)\}
  \nonumber\\
  &+\{\beta\hat p(k)+\alpha q(k+1)\}\lg\{\beta\hat
  p(k)+\alpha q(k+1)\}
  \nonumber\\
  &+\{\beta\hat p(k+1)+\alpha q(k)\}\lg\{\beta\hat
  p(k+1)+\alpha q(k)\}.
  \label{eq:theo:diffence}
 \end{align}
 Using the notation
 $a=\beta\hat p(k)+\alpha q(k)$,
 $b=\beta\hat p(k+1)+\alpha q(k)$, and
 $\varepsilon=\alpha q(k+1)-\alpha q(k)$,
 we have $a>b\ge0$ and $\varepsilon>0$.
 Now Eq.\,\ref{eq:theo:diffence} can be
 rewritten as
 \begin{align}
  &H(p^{\Pi^+})-H(\beta\hat p+\alpha q^{s_k})=\nonumber\\
  &-a\lg a -(b+\varepsilon)\lg(b+\varepsilon)
  +(a+\varepsilon)\lg(a+\varepsilon)+b\lg b
  > 0.\nonumber
 \end{align}
 where the last inequality holds from Lemma
 \ref{theo:lemm-convex}.
 This inequality indicates that
 $H(p^{\Pi^+})>H(\beta\hat p+\alpha q^{s_k})$.
 That is,
 \[
  [\pi_1,\cdots,\pi_{i-1}, s_k\circ\pi_i,\pi_{i+1},\cdots,\pi_\sigma]
 \]
 is better than $\Pi^+$.
 However, this contradicts
 the optimality of $\Pi^+$.
\end{IEEEproof}

\subsection{Proof of Theorem \ref{theo:proposedCB}}
\label{sec:CB-proof}
We can prove Theorem \ref{theo:proposedCB} in a
similar way to \cite{ImplicitCompBoost},
which proves the theorem for ICB with a balanced wavelet tree.
\begin{IEEEproof}
 To begin with, we introduce some facts
about RRR \cite{PracticalRRR}.
Let us consider a bit vector $B$ of length $n$.
The RRR divides $B$ into small blocks of length $b$:
$B=B^{(1)}B^{(2)}\cdots B^{(n/b)}$.
Each $B^{(j)}$ is represented by its \textit{class} $c_j$
and \textit{offset} $o_j$.
Here the class is the number of 1's in $B^{(j)}$,
and the offset is an index to distinguish the positions
of 1's in $B^{(j)}$.
In fact, the total space needed for the classes becomes the second term
of Eq.\,\ref{eq:rrr} (see \cite{PracticalRRR}).
Since this term is already considered in
Eq.\,\ref{eq:proposed_overhead}, what we have to
evaluate is the offsets.
Each offset requires $\lg\binom{b}{c_j}$ bits because
there are $\binom{b}{c_j}$ possible layouts of 1's for the class $c_j$.

Let us consider the partition of contexts of length $k$:
$\phi(T_{bwt})=L_1L_2\cdots L_l$ ($l\le\sigma^k$).
Since a bit vector in a node $v$ of a wavelet tree keeps the ordering,
the bit vector $B_v$ can be divided into $l$ blocks:
$B_v=B_1^vB_2^v\cdots B_l^v$.
Here each $B_j^v$ corresponds to $L_j$.

Now, we can consider small blocks of RRR which is fully included in
$B_j^v$.
Let us denote such blocks as $B^{(1)}B^{(2)}\cdots B^{(t)}$.
The offsets for these blocks require
\begin{align}
 \sum_{j=1}^{t}\lg\binom{b}{c_j}\le |B_j^v|H_0(B_j^v)
\end{align}
bits.
There are at most two blocks at the boundary of $B_j^v$
not considered above.
Their offset requires $O(b)$ bits;
therefore, the offsets for $B_j^v$ need $|B_j^v|H_0(B_j^v)+O(b)$ bits in total.

Let us consider the space needed for $L_j$.
Since there are at most $\sigma-1$ inner nodes in a wavelet tree,
summing the required spaces over $v$,
we obtain
\begin{align}
 \sum_v |B_j^v|H_0(B_j^v)+O(\sigma b) = |L_j|H_0(L_j)+O(\sigma b)
\end{align}
bits; the right hand side can be obtained by the recursive
calculation technique discussed in \cite{HuffWT-proof}.

Summing the above equation over $j$,
we have
\begin{align}
 \sum_{j=1}^l |L_j|H_0(L_j)+O(l\sigma b).
\end{align}
Although $L_j$ is an encoded string, the elements
have a one-to-one correspondence with the non-encoded string
because of the definition of the RML.
Hence, $H_0(L_j)$ is equal to $H_0(T_W)$, where
$W$ is the corresponding context and
$T_W$ is defined in Eq.\,\ref{eq:Hk}.
\end{IEEEproof}
\end{document}